\documentclass[preprint]{aastex}
\usepackage{color}
\usepackage{subfigure}
\usepackage[normalem]{ulem}
\usepackage{epsfig}
\usepackage{rotating}
\usepackage{lscape}
\usepackage{natbib}
\shorttitle{C/O Ratios in Transiting HJ Host Stars}
\shortauthors{Teske et al.}

\begin{document}

\title{C/O Ratios of Stars with Transiting Hot Jupiter Exoplanets {*,$^+$}}

\altaffiltext{*}{Based on data collected at Subaru Telescope, which is operated by the National Astronomical Observatory of Japan.} 
\altaffiltext{$^{+}$}{Some of the data presented herein were obtained at the W.M. Keck Observatory, which is operated as a scientific partnership among the California Institute of Technology, the University of California and the National Aeronautics and Space Administration. The Observatory was made possible by the generous financial support of the W.M. Keck Foundation.}

\author{Johanna K. Teske\altaffilmark{1}, Katia Cunha\altaffilmark{1, 2}, Verne V. Smith\altaffilmark{3}, Simon C. Schuler\altaffilmark{4}, Caitlin A. Griffith\altaffilmark{5}}

\altaffiltext{1}{Steward Observatory, University of Arizona, Tucson, AZ, 85721, USA; email: jteske@as.arizona.edu}
\altaffiltext{2}{Observat\'orio Nacional, Rua General Jos\'e Cristino, 77, 20921-400, S\~ao Crist\'ov\~ao, Rio de Janeiro, RJ, Brazil}
\altaffiltext{3}{National Optical Astronomy Observatory, 950 North Cherry Avenue, Tucson, AZ 85719, USA}
\altaffiltext{4}{University of Tampa, 401 W. Kennedy Blvd., Tampa, FL 33606, USA}
\altaffiltext{5}{Lunar and Planetary Laboratory, University of Arizona, Tucson, AZ, 85721, USA}

\begin{abstract}

The relative abundances of carbon and oxygen have long been recognized
as fundamental diagnostics of stellar chemical evolution. Now, the
growing number of exoplanet observations enable estimation
of these elements in exoplanetary atmospheres. In hot Jupiters, 
the C/O ratio affects the partitioning of carbon in the major
observable molecules, making these elements diagnostic of temperature structure and composition. Here we present measurements of carbon and oxygen
abundances in 16 stars that host transiting hot Jupiter exoplanets,
and compare our C/O ratios to those measured in larger samples of  
host stars, as well as those estimated for the corresponding 
exoplanet atmospheres. With standard stellar abundance analysis we derive 
stellar parameters as well as [C/H] and [O/H] from multiple
abundance indicators, including synthesis fitting of the [O I] 6300\,{\AA}
line and NLTE corrections for the O I triplet. Our results, in
agreement with recent suggestions, indicate that previously-measured
exoplanet host star C/O ratios may have been overestimated. The mean
transiting exoplanet host star C/O ratio from this sample is  0.54
(C/O$_{\odot}$=0.54), versus previously-measured
C/O$_{\rm{host~star}}$ means of $\sim$0.65-0.75. We also observe the
increase in C/O with [Fe/H] expected for all stars based on Galactic
chemical evolution; a linear fit to our results falls slightly below
that of other exoplanet host stars studies but has a similar
slope. Though the C/O ratios of even the most-observed 
exoplanets are still uncertain, the more precise abundance analysis
possible right now for their host stars can help constrain these planets' formation environments and current compositions.  

\end{abstract}


\clearpage

\section{Introduction}
 To date, the
most statistically significant trend in host star abundances pertains to metallicity.
Stars hosting giant, close-in planets have higher
metallicities (measured as [Fe/H]\footnote{[X/H]=log$N$(X) -
  log$N$(X)$_{\odot}$, where log$N$(X)=log$N$(X/H)+12}) than stars
without detected giant planets (e.g.,
Gonzalez 1998 \& 2001; 
Santos et al.\,2004; Fischer \& Valenti 2005; Ghezzi et al.\,2010). 
Statistical studies of dwarf stars hosting planets
indicate a metallicity
enhancement of $\sim$0.15 dex for stars with giant planets and a 99.9994\% 
probability that stars with/without giant
planets are drawn from 
different parent populations (Buchhave
et al. 2012; Ghezzi et al. 2010). However, the host
star metallicity trend is weaker for Neptune-sized planets -- the difference in the mean [Fe/H] of Jovian-mass hosts versus
Neptunian-mass hosts is $\sim$0.10 dex, with the Neptune-mass hosts
showing lower [Fe/H] values (e.g., Ghezzi et al.\,2010). Smaller planet
(R$_{\rm{P}}\leq$4 R$_{\oplus}$) host stars from the \textit{Kepler}
sample show
no metallicity enhancement, and have a flatter distribution of
metallicities, though roughly peaked at solar. These smaller planet
host stars have a
probability between 0.98 and 0.9996 of originating from a different parent population as
larger planet host stars from \textit{Kepler} (Buchhave et al.\,2012; Everett et al.\,2013). 

Looking beyond the correlation between planet size and stellar
metallicity, several studies have searched for other trends between
planet parameters and host star abundances indicative of planet
formation conditions. Mel\'endez et al.\,(2009) find through 
abundance analyses of 11 solar
``twins'' that the Sun is deficient by $\sim$20\% in refractory
elements, which have condensation temperatures $T_c\gtrsim$ 900 K, relative
to volatile elements when compared to other solar twins. 
This trend of decreasing refractory elemental abundances
as a function of $T_c$ is suggested to be a signature of terrestrial
planet formation -- the ``missing'' refractory elements from the
stellar photosphere are incorporated into rocky planets (Mel\'endez et
al.\,2009). However, subsequent studies of similar precision
measurements of solar analogs (Gonz\'alez Hern\'andez et
                                al.\,2010; Gonz\'alez Hern\'andez et
                                al.\,2013) and stars with planets
                                (Schuler et al.\,2011a \& 2011b) 
across a range of $T_c$ 
show that the abundance patterns of stars with and without
                                planets are not significantly different, or
                                may be indistinguishable from
                                Galactic chemical evolution
                                effects. New evidence from Jupiter- and Neptune-sized planet host stars
                                that are more metal-rich, or warmer
                                than the Sun and have less massive
                                convective envelopes, indicates that
                                the depletion signature may depend on the stellar convective envelope size at the
time of planet formation, and thus the timescale of disk dispersal around different types of stars (Ram\'{\i}rez et al.\,2013)

The growing number of transiting and directly-imaged exoplanet observations enable
estimates of elemental and molecular abundances in
the atmospheres of the planets themselves. The \textit{Hubble Space Telescope} and
\textit{Spitzer Space Telescope}, aided by multiple ground-based facilities, have detected the most abundant
molecules (H$_2$O, CO, CH$_4$, CO$_2$) in the atmospheres of several
of the brightest transiting planets (e.g., Tinetti et al.\, 2007;
Swain et al.\,2008, 2009ab; Snellen et al.\,2010; Beaulieu et al.\,2010). 
Ground-based observatories have made similar strides
in studying the molecular properties of a handful of directly imaged self-luminous 
exoplanets (e.g., Marois et al.\,2008; Barman et al.\,2011ab;
Skemer et al.\,2012, 2013; Konopacky et al.\,2013; Janson et
al.\,2013). The differences and trends between the elemental compositions of host
star and exoplanet atmospheres provide clues about the formation and
evolution processes of planetary systems.

\subsection{The Role of Carbon and Oxygen}
Carbon and oxygen are important players in the composition of stars
and planets, as the third
and fourth most abundant elements in the universe. 
The measurement of C and O in
stars, especially with respect to iron, which is produced in both
Type Ia and Type II supernovae, serves as a fundamental diagnostic of the
chemical enrichment history of the Galaxy. The impact of massive
stars' Type II supernovae, and thus the major oxygen contributor,
lessens with time and increasing metallicity as the influence of
low- and intermediate-mass stars' carbon contribution grows. Measuring
C and O in exoplanets is diagnostic of current
atmospheric composition and temperature structure:  
the atmospheric C/O ratio\footnote{The C/O ratio -- the ratio of carbon atom
   to oxygen atoms -- is calculated in stellar abundance analysis as C/O=
  N$_{\rm{C}}$/N$_{\rm{O}}$=10$^{\rm{logN(C)}}/10^{\rm{logN(O)}}$.}  
 affects the
molecular composition, and hence observed spectral signatures, through
thermochemial equilibrium partitioning of carbon in CO, CH$_{4}$, and
CO$_2$. 

The C/O ratio can also
reflect where in the protoplanetary disk a planet  
formed, as well as subsequent migration
and evolution (e.g., Stevenson \& Lunine 1988; Gaidos 2000; Ciesla \& Cuzzi
2006; \"Oberg et
al.\,2011).  
Theories of planet formation describe how close-in
giant planets 
form in the outer protoplanetary disk
, where icy planetesimals coalesce into a
core, which accretes gas and migrates inwards (e.g., Pollack et
al.\,1996; Owen et al.\,1999;  Ida \& Lin 2004b). The main molecular reservoirs of C and O
have different condensation temperatures ($T_c$), so their relative amounts vary at different
temperatures and disk radii, as do the amounts
of these molecules in gas or solid form (\"Oberg et al.\,2011). Gas and grains also move
differently in the disk with time, as grains grow and decouple from
the gas, sequestering solid material beyond the ``ice'' lines of
different molecules  (e.g., Ciesla \& Cuzzi 2006; \"Oberg et
al.\,2011). 
The C/O ratio
of a planet therefore does not necessarily reflect the
protoplanetary-disk-averaged C/O ratio, and instead may point towards
localized concentrations/depletions of carbon- and oxygen-bearing molecules 
(Ciesla \& Cuzzi 2006; \"Oberg et al.\,2011; Najita et al.\,2013). 
 
Many groups have performed stellar abundance analyses of exoplanet
host stars in order to determine their physical parameters
(T$_{\rm{eff}}$, log \textit{g}, [Fe/H]) and chemical abundances, 
to further study the trends discussed above (e.g., Delgado
Mena et al.\,2010; Petigura \& Marcy 2011; Brugamyer et al.\,2011; Schuler
et al.\,2011ab; Nissen 2013). However, only a few transiting
exoplanet host stars have published abundances other than [Fe/H] or the more generic
[M/H] (e.g., HD 209458, Schuler et
al.\,2011a; WASP-12, Petigura \& Marcy 2011; 55 Cnc, Bond
et al.\,2010 \& Teske et al.\,2013b; XO-2, Teske et al.\,2013a). 

Here we add to the small sample of 
transiting
exoplanet host stars with 
measured abundances beyond [Fe/H], and the handful with measured C/O
ratios. We 
report on sixteen transiting hot Jupiters hosts 
to investigate the extent to which we
can relate host star compositions to those of their planets, and
search for 
carbon-rich planet
formation environments. 
The sample presented here contains the host stars of some of the
most-observed exoplanets whose atmospheres can and are being modeled to
constrain their C/O ratios (e.g., Madhusudhan 2012; Moses et
al.\,2013; Line et al.\,2013). 
This work provides a step toward comparing specific host star and
exoplanet atmospheres to search for the 
chemical effects of exoplanet formation.

\section{Observations and Data Reduction}

Our target list was chosen to include some of the best-studied hot
Jupiter's host stars that are observable from the northern hemisphere,
and to include a range of planet radii, masses, and orbital periods. All
but three of the planetary hosts in this sample have at least the 3.6 $\mu$m, 4.5
$\mu$m, 5.8 $\mu$m, and 8.0 $\mu$m 
diagnostic
measurements of secondary eclipse depth  
from the \textit{Spitzer} Infrared
Array Camera (IRAC) (Fazio et al.\,2004). These data 
cover wavelengths
with features of CH$_4$, CO, CO$_2$, and H$_2$O, which are the most
abundanct oxygen and carbon molecules in hot Jupiter
atmospheres. These measurements, in addition to $HST$ and ground-based
observations, are analyzed to infer the carbon and oxygen
content in exoplanets (Moses et al. 2011, 2013; Lee et al.\,2012; Madhusudhan 2012; Line
et al. 2013), motivating our choice to target these host stars for C/O measurements.  
The planet orbiting HD 80606 has only 8.0 $\mu$m photometry, and the
planets orbiting HAT-P-16 and WASP-32 have only 3.6 $\mu$m
and 4.5 $\mu$m photometry, and CoRoT-2's planet is missing 5.8 $\mu$m photometry. We include these systems to increase the
planet mass range in this sample to include members of the 2.8\% of transiting
planets with masses M$\times$sin$i$ $>$
3 M$_{\rm{J}}$ -- HD 80606b is 3.94$\pm$0.11 M$_{\rm{J}}$
(Pont et al.\,2009), HAT-P-16b is 4.19$\pm$0.09 M$_{\rm{J}}$ (Buchhave et al.\,2010), and WASP-32b
is 3.60$\pm$0.07 M$_{\rm{J}}$ (Maxted et al.\,2010), and CoRoT-2b is
3.47 $\pm$0.22 M$_{\rm{J}}$
(Gillon et al.\,2010). 

There are three sources of observations for this project: the High
Dispersion Spectrograph (HDS; Noguchi et al.\,2002) on the 8.2-m
Subaru Telescope at Mauna Kea Observatory, the High Resolution Echelle
Spectrometer (HIRES; Vogt et al.\,1994) at the Keck I
Telescope, and the Keck/HIRES archive. All observations are logged
in Table \ref{tab:log}, and the platform configurations are detailed
in Table \ref{tab:platform}. 
To facilitate a differential abundance
analysis of these stars with respect to the Sun (as indicated by the
bracket notation), spectra of the Sun as reflected
moonlight were taken at Subaru/HDS, and as reflected light from Vesta
at Keck/HIRES (PID N014Hr; PI Marcy).

The HDS raw data were overscan-corrected, bias-subtracted,
scattered-light subtracted,
flat-fielded, extracted, and wavelength calibrated using standard techniques within the
IRAF\footnote{IRAF is distributed by the National Optical Astronomy
  Observatory, which is operated by the Association of Universities
  for Research in Astronomy, Inc., under cooperative agreement with
  the National Science Foundation.} software package using five bias
frames, 20 flat fields, and thorium argon (ThAr) comparison lamp
frames. All HIRES data were subject to a similar reduction procedure within the MAKEE
pipeline\footnote{www.astro.caltech.edu/~tb/makee/} using
corresponding bias ($\sim$3), flat ($\sim$30), ThAr, and trace
star frames.  
Multiple exposures of single targets were then summed in IRAF. 

We also obtained
Keck/HIRES archive spectra of four targets that we
originally observed with Subaru/HDS. The Subaru/HDS spectra were contaminated
by atmospheric emission around the [O I]
6300\,{\AA} line, preventing a secure measurement of [O/H]; the O I
triplet at $\sim$7775\,{\AA} was outside the Subaru/HDS
wavelength coverage. Thus the Keck/HIRES archival data were used to verify
the carbon abundance derived from the Subaru/HDS data, and to measure
the oxygen abundance as described below. The four targets for which
we obtained archival data are HAT-P-7 (PID
A285Hr; PI Bakos; August 2008), TrES-3 (PID C290Hr; PI Herczeg; June
2008), HD 189733 (PID A259Hr; PI Winn; August 2006), and HD 149026
(PID N59H; PI Marcy; June 2005).

\section{Derivation of Stellar Parameters and Abundances} 
We determine stellar parameters (T$_{\rm{eff}}$, log $g$,
microturbulence [$\xi$]) and elemental abundances of Fe, C, Ni,
and O following the procedures in Schuler et al.\,(2011a) and Teske et
al.\,(2013a). Briefly, the strength and shape of absorption lines in
stellar spectra depend on the formation environment (temperature,
electron pressure) and the number and excitations state of absorbers
themselves and thus their atomic constants. Thus one uses
measurements from abundant, unblended lines in multiple ionization
states -- typically Fe -- to determine the stellar
environment in which the observed line strengths form, in an interative
manner. The ``best'' stellar model parameters -- effective temperature,
microturbulent velocity, log $g$, and [Fe/H] -- results from fulfilling
excitation equilibrium such that the [Fe/H] values
derived from the Fe I lines do not show any correlation with the
lower excitation potential of the lines ($\chi$), ionization
equilibrium such that averaged abundances from Fe I and Fe II lines
are equal, and ensuring that Fe I lines of all different equilvalent widths yield consisten abundances. 

Specifically, the 
Fe
lines in this analysis are the same as in Teske et al.\,(2013ab).  
The final Fe line list contains 56 Fe I and 10 Fe II lines, although
not every Fe line is measureable in every star in the sample. Lower
excitation potentials 
and transition probabilities 
for the Fe lines are from the Vienna Atomic Line Database (VALD; Piskunov et
al.\,1995; Kupka et al.\,1999; Ryabchikova et al.\,1999). 

\begin{figure}[ht!]
\figurenum{1}
\centering
\includegraphics[width=.75\textwidth]{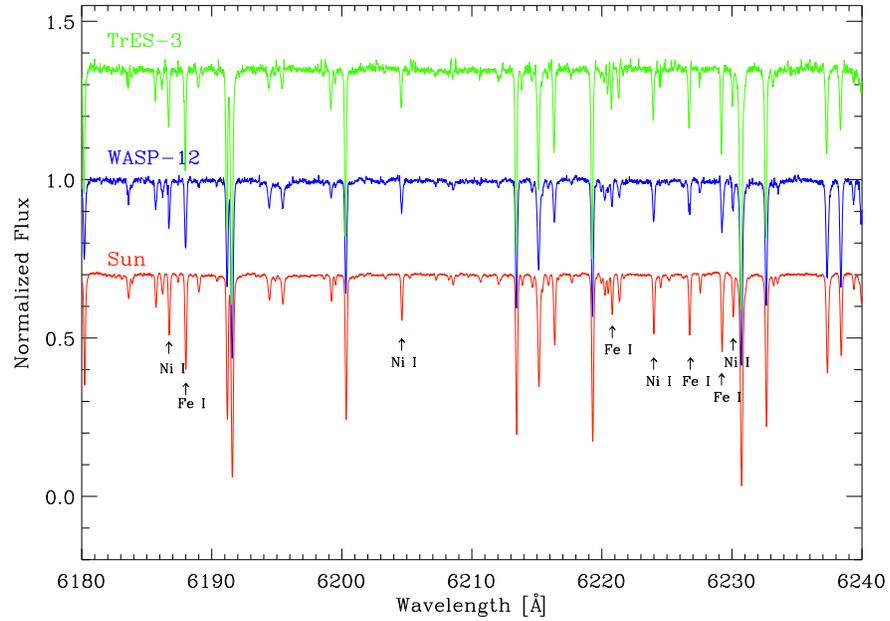}
\caption{Sample spectra of TrES-3 (green), WASP-12 (blue), and the Sun
  (red) obtained with Subaru/HDS. The spectra have been continuum normalized and are
  shifted by constant values in flux for ease of viewing. Lines in
  this order for which EWs were measured are marked with arrows.}
\label{spectra_plots}
\end{figure}

\begin{figure}[ht!]
\figurenum{2}
\centering
\includegraphics[width=.75\textwidth]{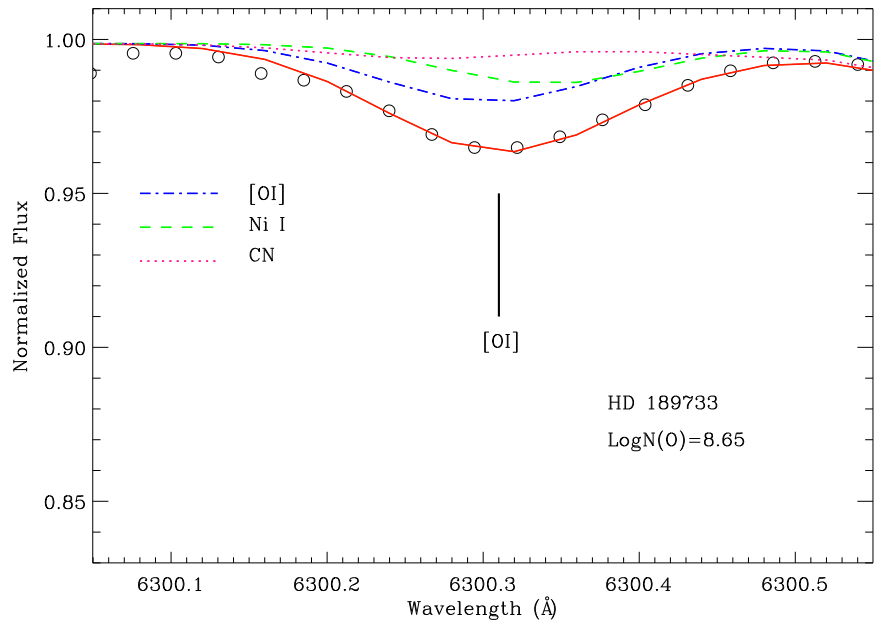}
\caption{Shown here is the spectrum synthesis fit to the forbidden
  [OI] line (6300.3 \{AA) for HD 189733. The data are shown as black
  open circles. The full synthesis fit is represented by a solid red
  line, with components shown with blue dash-dotted ([OI]), green
  dashed (Ni I), and pink dotted (CN) lines.}
\label{synth_fig}
\end{figure}

Abundances of Fe, C, and Ni 
 are derived 
from equivalent width (EW) measurements of spectral lines in each
target individually (sample Subaru/HDS spectra are shown in
Fig.\,\ref{spectra_plots}). The EW measurements are performed with
the goal of mitigating errors of the fit -- we use
a Gaussian profile to fit most lines, though some strong lines (EW$\geq$90
m{\AA}) are fit with a Voigt profile to account for the broader wings
of the line at the continuum, and some weaker lines are fit with a
Simpson's Rule approximation. The lines of the host stars are fit with
either the one-dimensional spectrum analysis package SPECTRE
(Fitzpatrick \& Sneden 1987) or the `splot' task in
IRAF. The solar spectra corresponding to each target used to derive
differential abundances were fit with the same package as the target, e.g., where we use EW measurements from SPECTRE, we use corresponding EW
measurements from SPECTRE of the Sun in our analysis. 
The abundances are then determined with an updated version of the local
thermodynamic equilibrium (LTE) spectral analysis code MOOG (Sneden
1973), with model atmospheres interpolated from the Kurucz ATLAS9
grids\footnote{See http://kurucz.harvard.edu/grids.html}. 
For oxygen, we use the spectral synthesis method of matching a set of trial synthetic spectra to
the observed spectrum 
derive the abundance from the blended [O I] line at $\lambda$ = 6300.3\,{\AA} (see
Fig.\,\ref{synth_fig}). 

Initial values of T$_{\rm{eff}}$, log $g$, $\xi$,
and [Fe/H] from the literature 
serve as our starting values in
the iterative process of meeting the criteria outlined
above. 
Prior to this
iterative scheme, for each target's EW measurements we 
ensure no correlation between $\chi$ and the EWs of the Fe I
lines analyzed, as unique solutions for T$_{\rm{eff}}$ and $\xi$ are only
possible if there is no such initial correlation. 
The log$N$(Fe) values for
the Sun are determined from the solar spectrum 
with a solar Kurucz model with T$_{\rm{eff}}$=5777, log $g$=4.4
[Fe/H]=0.00, and $\xi$=1.38, and log$N$(Fe) values of the target stars 
are normalized to solar values on a line-by-line basis. 
The final [Fe/H] results from averaging
the abundances derived from the individual Fe I and Fe II lines. 

Uncertanties in the derived stellar parameters are calculated as
detailed in Teske et al.\,(2013a). In Table \ref{tab:stellar_params} we list the final derived stellar
parameters and their 1$\sigma$ uncertainties for each target, as well
as the derived [Fe I/H] and [Fe II/H], the number of lines
used in our analysis, and the uncertainy in the mean ($\sigma_{\mu}$\footnote{$\sigma_{\mu}=\sigma/\sqrt{N-1}$,
where $\sigma$ is the standard deviation of the derived abundances and
$N$ is the number of lines used to derive the abundance.}). 

\subsection{Stellar Abundances of Ni, C, and O}
A similar procedure as that used for the Fe lines is used to identify
and select lines for Ni and C, with the same line lists as Teske et
al.\,(2013ab), 
which include 5 lines for carbon and 20 lines for nickel, though not
every line was measureable in every star in our sample. 
The C and Ni abundances are
determined through standard EW analysis procedures 
with MOOG and the adopted stellar model for each target. The wavelength, $\chi$, log $gf$, measured EW, and resulting
abundances for the 
carbon lines  
are listed in Table
\ref{tab:lines} for a sample of the targets. All lines parameters,
equivalent widths, and resulting abundances are available in the
full online version of Table \ref{tab:lines}. 

We favor the two bluest C I lines (5052\,{\AA} and 5380\,{\AA}) in our
analysis because they arise from the lowest energy levels
considered here and have negligible non-LTE (NLTE) corrections ($\leq$0.05; Asplund et al.\,2005; Takeda \& Honda 2005; Caffau et
al.\,2010). The log$N$(C)$_{\odot}$ values we derive with our EW measurements are
a good match (within $\leq$0.03 dex) to the
log$N$(C)$_{\odot}$ values derived by Caffau et al.\,(2010) from these
lines using 3D hydrodynamical simulations of the Sun. 
The remaining
three C I lines arise from higher energy levels, potentially making
them more susceptible to NLTE effects (Asplund 2005), although Asplund
et al.\,(2005) find NLTE corrections for these lines in the Sun are
comparable to those for the bluer C I lines. In cases where we find discrepant (larger) carbon abundances from
the redder C I lines, we
base our [C/H] measurement on the two bluest C I lines. 

In HAT-P-7, TrES-3, HD 149026, and HD 189733, [O/H] is not measurable
with the Subaru/HDS data because the [O I] 6300\,{\AA} is 
contaminated by atmospheric emission and the wavelength coverage ends blueward of the O I
triplet. The oxygen abundances of these four targets are instead measured from Keck/HIRES archive data (detailed below). In these cases the stellar
parameters and [Ni/H] derived from the Subaru/HDS data are
carried through the analysis, but [C/H] is remeasured in the
Keck/HIRES data so that the C/O ratio originates completely from
one data source. In all four cases the [C/H]$_{\rm{HIRES}}$ value is equal
to the [C/H]$_{\rm{HDS}}$ value within errors, with the differences
all $\leq$0.04 dex, giving confidence to this method.  

\subsubsection{Oxygen}
The oxygen abundances of the targets in our sample are determined from
two indicators, the forbidden [O I] line at $\lambda$ = 6300.3\,{\AA}
and the O I triplet at 7771-7775\,{\AA}, depending on what data are
available. For seven of the thirteen (including XO-2S) targets, the
oxygen abundance is derived solely from the Subaru/HDS data and the [O
I] line,  
which is well-described by LTE
(e.g., Takeda 2003). We adopt the Storey \& Zeippen (2000) log $gf$ =-9.717 value, based
on their forbidden transition probability calculations including both
relativistically-corrected magnetic dipole and electric quadruopole
contributions. However, the 6300.3\,{\AA} feature is blended with 
a Ni I line
composed of two isotopic components, with log $gf$($^{60}$Ni)=-2.695 and log
$gf$($^{58}$Ni)=-2.275 (Johansson et al.\,2003; Bensby et al.\,2004). In the Sun, nickel
accounts for $\sim$30-40\% of the [O I] absorption line depth (Caffau et al.\,2008; 2013). Therefore, in determining [O/H] from this line, we account for the nickel
component by measuring [Ni/H]
directly from each target's spectrum as described above and
appropriately scaling the strength of the blend component due to
nickel. We also test for potential
  blending in the 6300.3\,{\AA} [O I] line with another, 
much weaker, CN line (Davis \& Phillips 1963; Sneden \& Lambert 1982) by fixing the carbon abundance to our measured
[C/H] value and remeasuring the 6300\,{\AA} line oxygen abundance. Except in the cases
of XO-2N and XO-2S (Teske et al.\,2013b), our resulting oxygen abundances do not change
significantly ($\leq$0.02 log$N$(O)) by fixing the carbon abundance.

The free parameters in our synthesis fits are the continuum normalization, wavelength shift, line broadening, and oxygen abundance; we use 
our measured Ni and C abundances for each star, and N scaled from
solar based on
the measured [Fe/H] of each star. We also checked our synthesis
results with the ``blends'' driver in MOOG, which accounts for
contributions from additional  
lines to the primary element with a user-provided
line list including these blending lines and corresponding blending
species abundances. 
In Table \ref{tab:lines} we
list the measured EWs serving as input for our check with the
``blends'' driver and absolute abundances as determined from
synthesis fitting of the [O I] line for a
sample of the targets. 

For the spectra in which the Subaru/HDS [O I]
6300.3\,{\AA} is tellurically contaminated (HAT-P-7, TrES-3, HD 149026, and HD 189733)
 the even weaker [O I] 6363.79\,{\AA}
line is not distinguishable from noise, and the O I triplet at
7771-7775\,{\AA} is not covered. 
However, the public data retrieved from the Keck/HIRES
archive do include the triplet at 7771-7775\,{\AA} and, in some cases,
also display clean (not tellurically contaminated) [O I] 6300\,{\AA}
lines. In TrES-3, the data beyond $\sim$7050\,{\AA} are contaminated
due to saturated ThAr lamp calibrations; fortunately in this case the [O I] 6300\,{\AA} is
measurable in the Keck/HIRES data. We also obtained our own
Keck/HIRES data for HAT-P-1, HAT-P-16, and WASP-32, and use the
[O I] 6300\,{\AA} and O I triplet lines to derive oxygen abundances for
these targets, as well as the stellar parameters and the [C/H] and
[Ni/H] values.  

In solar-type stars, the O I triplet lines at 7771.94\,{\AA}
($\chi$=9.15 eV, log $gf$=0.369; Hibbert et al.\,1991), 7774\,{\AA}
($\chi$=9.15 eV, log $gf$=0.223; Hibbert et al.\,1991), and
7775.4\,{\AA} ($\chi$=9.15 eV,log $gf$= 0.001; Hibbert et al.\,1991) 
are prominent and suffer less from blending with other lines, and are therefore conducive to direct EW measurement (Table
\ref{tab:lines}). 
These lines are known to suffer from NLTE
effects, detailed in Kiselman (1993; 2001) and Gratton et al.\,(1999),
and several groups have derived NLTE corrections  
from statistical
equilibrium calculations
for varying stellar parameters. For comparison, as in Teske et al.\,(2013a),
we apply  
NLTE corrections to the triplet abundances from three sources: Takeda (2003), Ram{\'{\i}}rez et
al.\,(2007), and Fabbian et al.\,(2009) (Table
\ref{tab:oh}). The differences in methodology applied by each
of these sources to determine NLTE corrections is detailed in Teske et
al.\,(2013a), who also note that the NLTE corrections from different
sources give overlapping results.

The validity of using such NLTE corrections for cool
(T$_{\rm{eff}}\lesssim$5400 K) stars is questionable, as discussed in Teske et al.\,(2013a). 
They ultimately choose not to include the NLTE-corrected O I triplet
abundances for the cool, metal-rich star 55 Cnc, instead relying on
the averaged [O I] 6363.79\,{\AA} and LTE O I triplet abundances. In the
spectra presented here, which are of lower S/N than the 55 Cnc data presented in Teske et al.\,(2013a), the [O I] 6363.79\,{\AA} line is not
detected. Thus, in the case of stars with T$_{\rm{eff}}\lesssim$5400 K
and for which we have Keck/HIRES data with
wavelength coverage including the O I triplet, we
adopt the average of the [O I] 6300\,{\AA} and LTE O I triplet
abundances. For warmer stars in which both the [O I] 6300\,{\AA} and
O I triplet lines are measurable, we adopt the average of the [O I]
6300\,{\AA} and the three NLTE-corrected O I triplet abundances. 
In all cases the [O I] 6300\,{\AA}- and O I triplet-derived
(whether LTE or NLTE) abundances
match  
within the uncertainties.

Table \ref{tab:compare} lists the final averaged [O/H] values for each
target, along with [C/H] and the resulting C/O ratio. 
These C/O ratios are calculated with the prescription
log$N_{\rm{target}}$(O)$=$derived\,[O/H]$_{\rm{target}}$+log$N_{\odot}$(O)
and log$N_{\rm{target}}$(C)$=$derived\,[C/H]$_{\rm{target}}$+log$N_{\odot}$(C),
where log$N_{\odot}$(O)$=$8.66 and log$N_{\odot}$(C)$=$8.39 (solar
values from Asplund et al.\,2005). 
This
table also shows the measured [Ni/H] abundances for each target, which
are used in the derivation of [O/H] from the [O I] 6300\,{\AA} line as described above. 

\subsection{Abundance Uncertanties}
The uncertainties in the derived elemental
abundances include the errors in the adopted stellar parameters
(T$_{\rm{eff}}$, log $g$, and $\xi$) and the dispersion in the
abundances derived from different absorption lines for the element, as
the final adopted abundance is an average of these lines. To
determine the uncertainty due to the stellar parameters, the sensitivity
of the abundance to each parameter was calculated for changes of
$\pm150$ K in T$_{\rm{eff}}$, $\pm0.25$ dex in log $g$, and
$\pm0.30$km s$^{-1}$ in $\xi$. These calculated abundance
sensitivities for two targets, WASP-12 and HAT-P-1, are shown as an
example in Table \ref{tab:sens}. The final uncertainty due to each
parameter is then the product of this sensitivity and the
corresponding scaled parameter uncertainty, as described in Teske et
al.\,2013a. The dispersion in the abundances derived from
different lines is parameterized with the uncertainty in the mean,
$\sigma_{\mu}$
, for the abundances derived from the averaging of
multiple lines. Then the total internal uncertainty for each abundance
($\sigma_{\rm{tot}}$) is the quadratic sum of the individual parameter
uncertainties 
and $\sigma_{\mu}$. 

In the case of the O I triplet, the error on [O/H]$_{\rm{NLTE}}$ was
calculated separately for each of the applied NLTE corrections, but as
these errors are smaller than the error derived from the LTE
measurement, the LTE errors are conservatively adopted. In the cases of more than one
measurable oxygen abundance indicator, the errors associated with
[O/H] reported in Table \ref{tab:compare} are the errors from each
oxygen abundance indicator ([O I] 6300\,{\AA} and LTE O I triplet) added in quadrature, unless otherwise noted. Similarly, the C/O ratio errors are the errors of [C/H] and [O/H] combined in quadrature. 

\section{Results and Discussion}
Our final adopted stellar parameters and
their 1$\sigma$ uncertainties for each target are listed in
Table \ref{tab:stellar_params}, and the adopted elemental abundances
and their 1$\sigma$ uncertanties for each target are listed in Table
\ref{tab:compare}. 
We compared our results to those of the catalog of Stars With ExoplanETs (SWEET-Cat), described in Santos et al.\,(2013),
and those determined by Torres et al.\,(2012; T12). 
These two references are
comparable to ours in their analysis methods and samples (transiting
exoplanet host stars). SWEET-Cat compiles sets of atmospheric
parameters previously published in the literature and, whenever
possible, derived using the same uniform methodology of Santos et
al.\,(2004). The main sources of stellar  
parameters in SWEET-Cat for
the targets in our sample are Santos et al.\,(2004), Ammler-von Eiff et al.\,(2009),
and Mortier et al.\,(2013). SWEET-Cat reports T$_{\rm{eff}}$, [Fe/H],
and log $g$ for all of the targets in our sample of stars, and $\xi$ for ten of the
targets. 

T12 compared the
resulting T$_{\rm{eff}}$ and [Fe/H] from three different
stellar analysis programs. They include stellar parameter classification (SPC;
Buchhave et al.\,2012), Spectroscopy Made Easy (SME; Valenti \&
Piskunov 1996), and MOOG, the latter of which we employ here. They
also determine log $g$ values, but do not report them, and $v$ sin $i$
values, which we do not determine in our work. T12 report
final ``averaged'' T$_{\rm{eff}}$ and [Fe/H] values from
all attempted analysis methods for thirteen of the stars in our sample, though 
MOOG-derived T$_{\rm{eff}}$ and [Fe/H] values are included in that
average for only ten of the stars in our sample.

In all cases our derived
[Fe/H] values are consistent with those of SWEET-Cat and
T12-average, within uncertainties. The median $\Delta$[Fe/H] (as in
$|$ours - theirs$|$) values are 0.08 and 0.02 for SWEET-Cat and T12-average,
respectively. The
T$_{\rm{eff}}$ and log $g$ values reported here also overlap within uncertainties the
values reported in T12-average and those in SWEET-Cat in almost every
case. The median
$\Delta$log $g$ for SWEET-Cat is 0.17, and the median $\Delta$T$_{\rm{eff}}$
for SWEET-Cat and T12-average are 55 K and 41 K, respectively. In the two cases where
T$_{\rm{eff}}$ does not overlap SWEET-Cat our values are cooler by 54 K
(WASP-12) and 202 K (WASP-32). Two other sources (Brown et al.\,2012; Maxted et al.\,2010) report a very similar
T$_{\rm{eff}}$ for WASP-32 as the (cooler) value derived here, and the
results of Torres et al.\,(2012) for WASP-12 agree well with our
T$_{\rm{eff}}$. In the three cases where log $g$ does not overlap
SWEET-Cat or T12-average our values differ 
by $\leq$0.04 dex. 

\subsection{Comparison to Previous Studies of C and O in Exoplanet Host Stars}
This study focuses on transiting exoplanet host star 
elemental abundances, particularly their C/O ratios. No previous study
of which we are aware has uniformly derived [C/H], [O/H], and C/O
values for these stars. 
However, several studies have examined C/O ratios in non-transiting
exoplanet host stars versus  
stars not known to host planets. [Any star
designated as a ``non-host'' has the potential to 
harbor a smaller ( 
undetected) planet; indeed, it may be the case that most stars have
one or more small planets (e.g., Cassan et al.\,2012).] Here we
compare our results to these other host star C/O studies.

Bond et al.\,(2006) measure [C/H] in 136 G-type stars, 20
of which are exoplanet hosts, and Bond et al.\,(2008) measure [O/H]
ratios in 118 F- and G-type stars stars, 27 of which are known exoplanet hosts. Line
lists are not explicitly given for the measured C lines in Bond et
al.\,(2006); Bond et al.\,(2008) use the high-excitation O I triplet
at $\lambda$ = 7771.9, 7774.2, and 7775.4\,{\AA}. 
Bond et al.\,(2010) also compiled 
C/O ratios derived from measurements in Ecuvillion et al.\,(2004) and (2006). 
Ecuvillion et al.\,(2004) measures [C/H] from
the two lowest excitation lines of carbon
(5052.17 and 5380.34\,{\AA}), and 
Ecuvillion et al.\,(2006) measures [O/H] from the forbidden [O I] line at 6300.3
\AA, the high-excitation O I triplet at $\lambda$ = 7774 
\AA, and a set of 5 near-UV OH lines around 3100
\AA. Both of the Ecuvillion studies and the Bond studies implement an
analysis method similar to
that performed here, with the spectral synthesis code MOOG and a grid
of Kurucz (1993) ATLAS9 model atmospheres, although these studies do
not derive the host star parameters (T$_{\rm{eff}}$, log \textit{g},
$\xi$, and [Fe/H]), only specific elemental abundance ratios. In the
host star sample reported in Bond et al.\,(2010), 35\% have C/O$>$0.8
(they did not specifically report any non-host stars); Bond et al.\,(2008)
find 8\% of host stars and 5\% of non-host stars in their sample to have C/O$>$0.8. 

Following the Bond et al. investigation, two larger studies of the C/O
ratios of non-host stars versus host stars were conducted. 
Delgado Mena et al.\,(2010) measure carbon and oxygen in 100 host
stars,  
along with 270 non-host
stars, using  the C I lines at 5052.17 {\AA} and 5380.34\,{\AA} and the
[O I] forbidden line at 6300.3 {\AA}. They measure equivalent widths with
the ARES program\footnote{The ARES code can be downloaded at
  http://www.astro.up.pt/∼sousasag/ares/.} (Sousa et al.\,2007), and
used 
MOOG 
and Kurucz ATLAS9 model atmospheres (Kurucz\,1996)
for abundance analysis. Delgado Mena et al.\,(2010) find 34\% of their measured
host stars have C/O$>$0.8, while in
their non-host sample the fraction of stars with C/O$>$0.8 is 20\%.
Petigura \& Marcy\,(2011) find carbon and oxygen abundances for 704
and 604 stars, respectively, but only 457 have reliable measurements
for both elements that can be used to determine C/O ratios, 99 of
which are exoplanet hosts. These authors measure the 6587 {\AA} C I
line for carbon and the [O I] line at 6300.3 {\AA} for oxygen, and use
the SME code  
with
Kurucz (1992) stellar atmospheres for their abundance analysis. Petigura \& Marcy\,(2011) find 
34\% of host stars in their sample have C/O$>$0.8, versus 27\% of
non-host stars in their sample with with
C/O$>$0.8. 

The goal of this paper is to investiage and constrain values of
stellar host C/O ratios in systems with observed transiting giant
planets, since transit spectroscopy potentially allows for
determinations of the corresponding planetary C/O ratios. This goal is
driven partially by the recent suggestion by 
Fortney\,(2012) that the C/O ratios of both host- and non-host
stars in the studies noted above have been overestimated due to errors in the derived
C/O ratios and the observed apparent 
frequency of carbon dwarf
stars implied by these studies. 

Nissen\,(2013) recently rederived the carbon and oxygen
abundances for 33 of Delgado Mena et
  al.\,(2010)'s host stars that have additional ESO 2.2m FEROS spectra
  covering the O I triplet at 7774\,{\AA}, which was not originally used by
  Delgado Mena. 
He implements a differential analysis
with respect to the Sun, with equivalent widths of C and O
measured in IRAF with Gaussian profiles and abundances derived by
matching the observed equivalent widths with those measured in
plane parallel MARCS atmosphere models (Gustafsson et al.\,2008)
having the same stellar parameters as those  published by
Delgado Mena et al.\,(2010). Accounting for NLTE effects on the
  triplet line strengths by using the Fabbian et al.\,(2009)
  corrections, Nissen\,(2013) finds  
differences from Delgado Mena et
  al.\,(2010) in the
  derived oxygen abundances. 
This results in  
both a 
tighter correlation between [Fe/H] and C/O (Nissen finds C/O = 0.56$+$0.54[Fe/H] with
  an rms dispersion $\sigma$(C/O)=0.06), as well as a much smaller
  fraction of host stars with 
C/O$>$0.8 (only 1 out of 33).

The tight trend of increasing C/O with [Fe/H], e.g., Nissen (2013; as
noted above), is indicative of the importance of overall Galactic
chemical evolution in setting the fraction of dwarf stars that might
be carbon rich. The increase in C/O with metallicity points to the
importance of low- and intermediate-mass star carbon nucleosynthesis
at later, more metal-rich times. The influence of mass-star Type II
supernovae, the major oxygen contributors, is diluted with time as
low- and intermediate-mass stars become more important, thus C/O
increases. The fraction of ``carbon-rich'' (C/O $\geq$0.8)
planet-hosting stars is thus expected to increase with increasing
metallicity in the disk, with the Nissen trend indicating
metallicities greater than [Fe/H]$\sim$0.4 might begin to have
significant fractions of carbon-rich dwarf stars. All of the
planet-hosting stellar samples discussed here have very few, if any,
stars at these metallicites or higher. 

\begin{figure}[htb]
\figurenum{3}

   \subfigure{\includegraphics[width=0.5\textwidth]{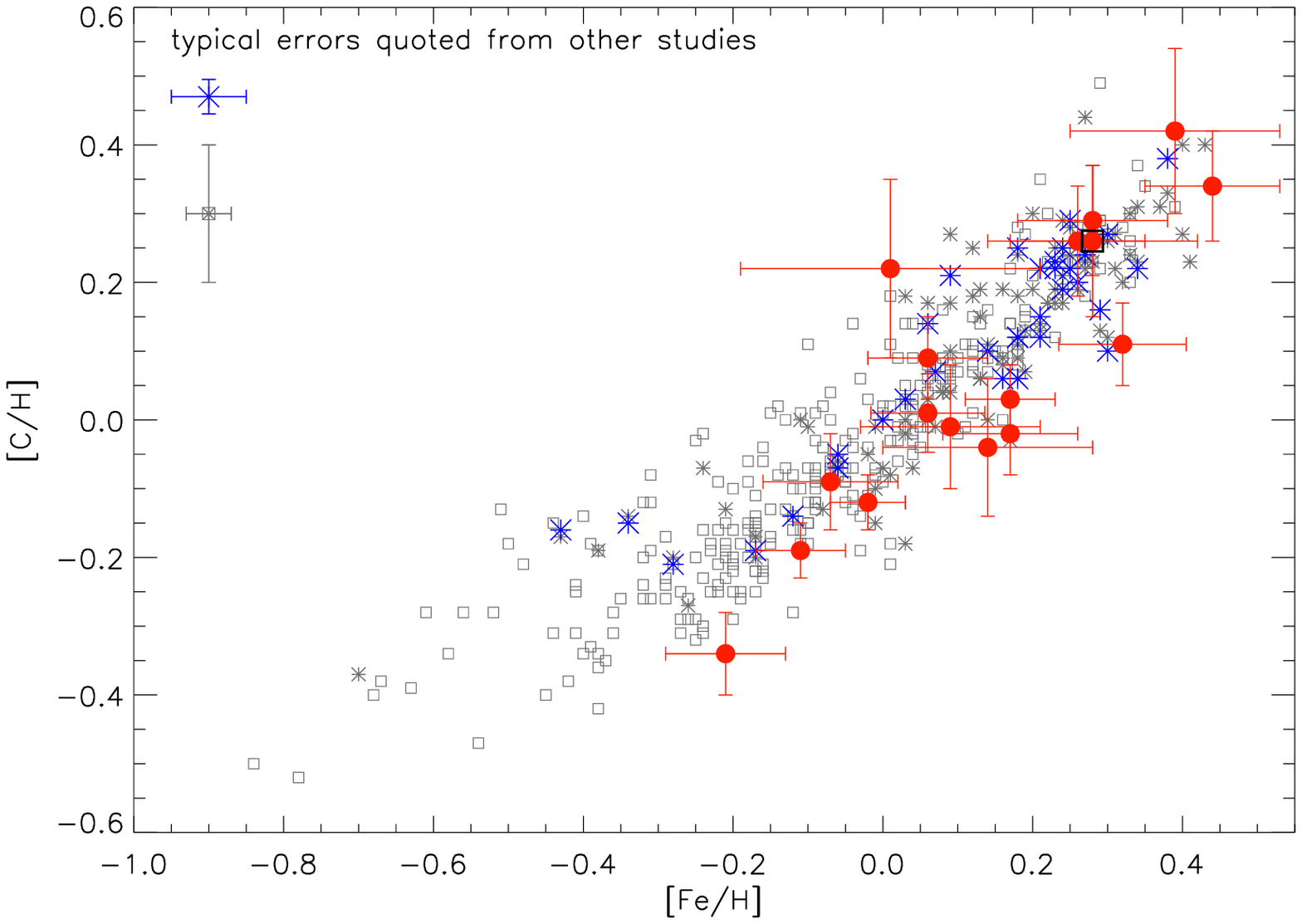}}
\quad
   \subfigure{\includegraphics[width=0.5\textwidth]{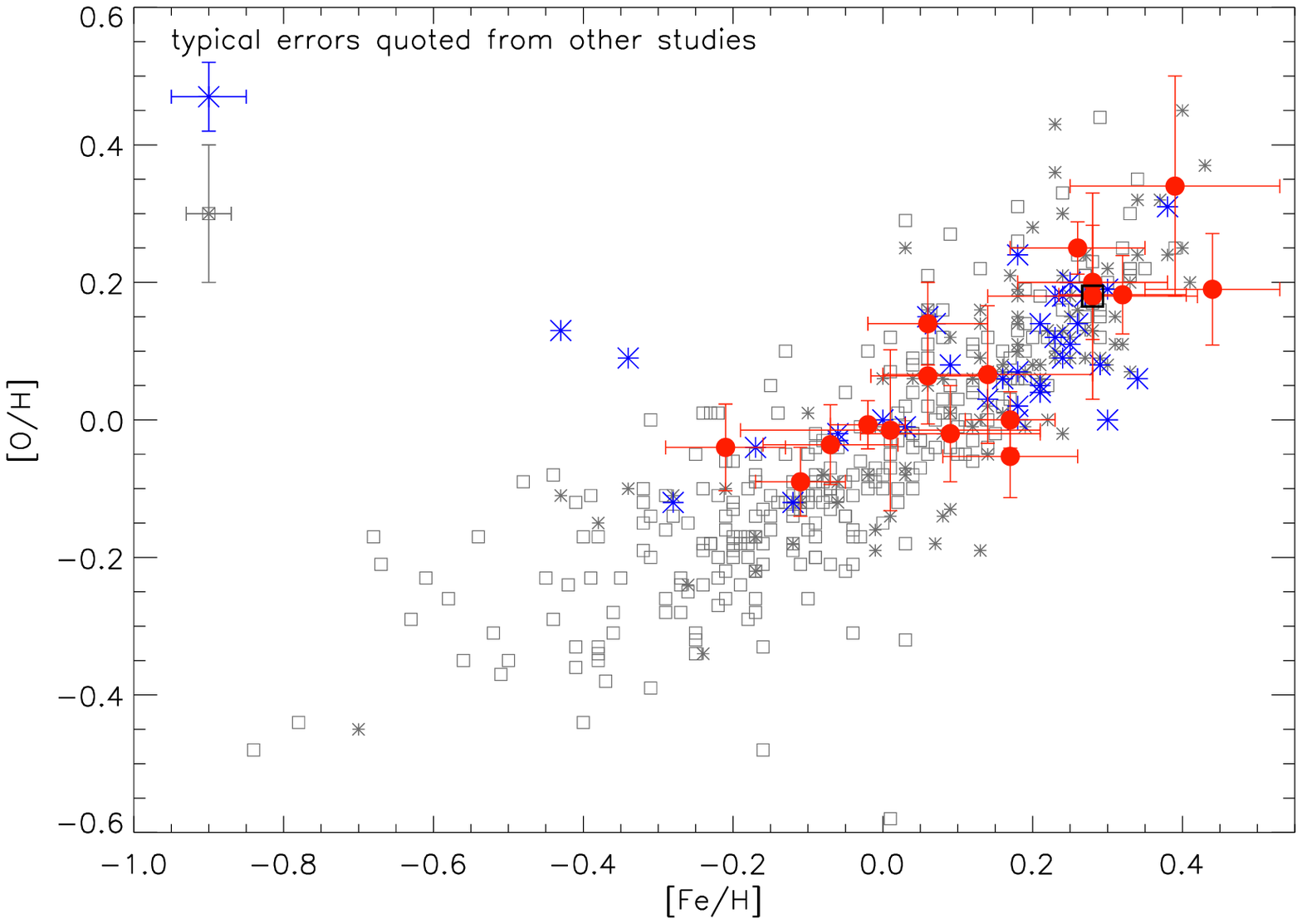}}
\caption{[C/H] and [O/H] versus 
  [Fe/H] from Delgado Mena et al.\,(2010) and Nissen\,(2013) [all
  Nissen\,(2013) hosts are in the Delgado Mena et al.\,(2010) host
  sample]. Non-host stars from Delgado Mena et al.\,(2010) are plotted
  with gray open squares, while host stars from Delgado Mena et
  al.\,(2010)/Nissen\,(2013) are plotted with gray/blue
  asterisks. Quoted typical error bars are in the
  upper left. Measurements from this work are plotted as red filled
  circles, with error bars included (see Table
  \ref{tab:compare}). In particular, XO-2S is plotted as a red circle enclosed
  by a black square, to indicate that it does not host a known planet (in
  the C/O plot, XO-2N overlaps XO-2S).}
\label{comparison_plot1}
\end{figure}

This paper differs from the studies listed above because 1) the
sample here is much smaller, being limited to only hosts of transiting 
exoplanets, and 2) only one non-host star is included (the binary
companion XO-2S). In Figure \ref{comparison_plot1} the host star
abundances derived here, shown with red filled circles with error bars, for [C/H] and [O/H] versus [Fe/H] are 
compared to the results of the large
samples of Delgado Mena et
al.\,(2010), shown as gray asterisks for host stars and open squares for
non-host stars, and Nissen\,(2013), shown as blue asterisks. All three studies define similar
behaviors of [C/H] and [O/H] as a function of [Fe/H], with the carbon
exhibiting larger slopes with iron relative to oxygen; this
illustrates the increasing importance of carbon production from low-
and intermediate-mass stars relative to massive stars with increasing
chemical maturity. 

The slopes of the trends in Figure \ref{comparison_plot1} are all
quite similar, with the Nissen (2013) trend exhibiting the smallest
scatter about a linear fit. Using the results of this paper, linear trends are
fit to both [C/H] and [O/H] versus [Fe/H] with the following results:
[C/H] = 0.95[Fe/H] - 0.05 and [O/H] = 0.56[Fe/H] + 0.01. Excluding the
apparently carbon-rich outlier HD 189733, these fits
are [C/H] = 1.02[Fe/H] - 0.08 and [O/H] = 0.56[Fe/H] +
0.01; we refer to these fits without HD 189733 throughout the rest of the
paper. Quantitatively, the C versus Fe slope is about twice as large
(in dex) as that for O versus Fe based on the linear fits to the
abundances derived here. The relation for carbon passes 
0.08 dex below solar (i.e., at [Fe/H],[C/H]=0,0)
while passing close to solar for oxygen, offset by only +0.01 dex.

Because of these even rather small offsets
($\sim$0.05-0.1 dex), the C/O ratios as a function of [Fe/H] might
fall below solar as defined by our results for this particular sample
of stars: an offset of -0.05 to -0.1 dex in [C/O] would correspond to
an offset of 0.1 to 0.2 lower in a linear value of C/O. This does not
necessarily correspond to simply errors in the analysis, but may
reflect both fitting linear relations to our results, which are
probably only approximate descriptions of the real Galactic disk relation, as
well as there not being a universal trend of [C or O]/H versus
[Fe/H]. The offsets here most probably reflect both uncertainties in
the analysis (already discussed in Section 3.2) and intrinsic scatter
in real Galactic disk populations that will map onto the sample
of stars analyzed here. 

\begin{figure}[htb]
\figurenum{4}
   \subfigure{\includegraphics[width=0.5\textwidth]{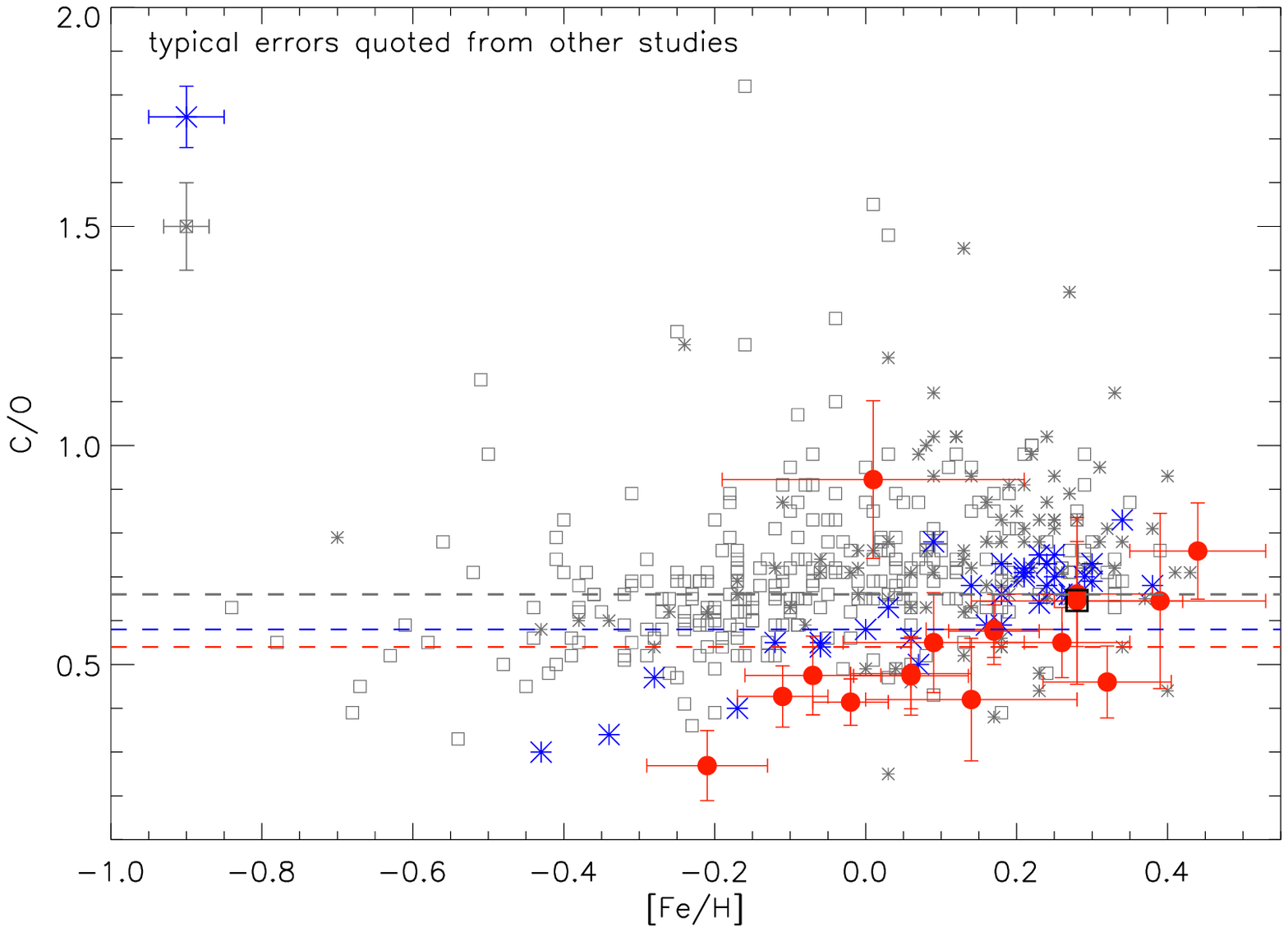}}
\quad
   \subfigure{\includegraphics[width=0.5\textwidth]{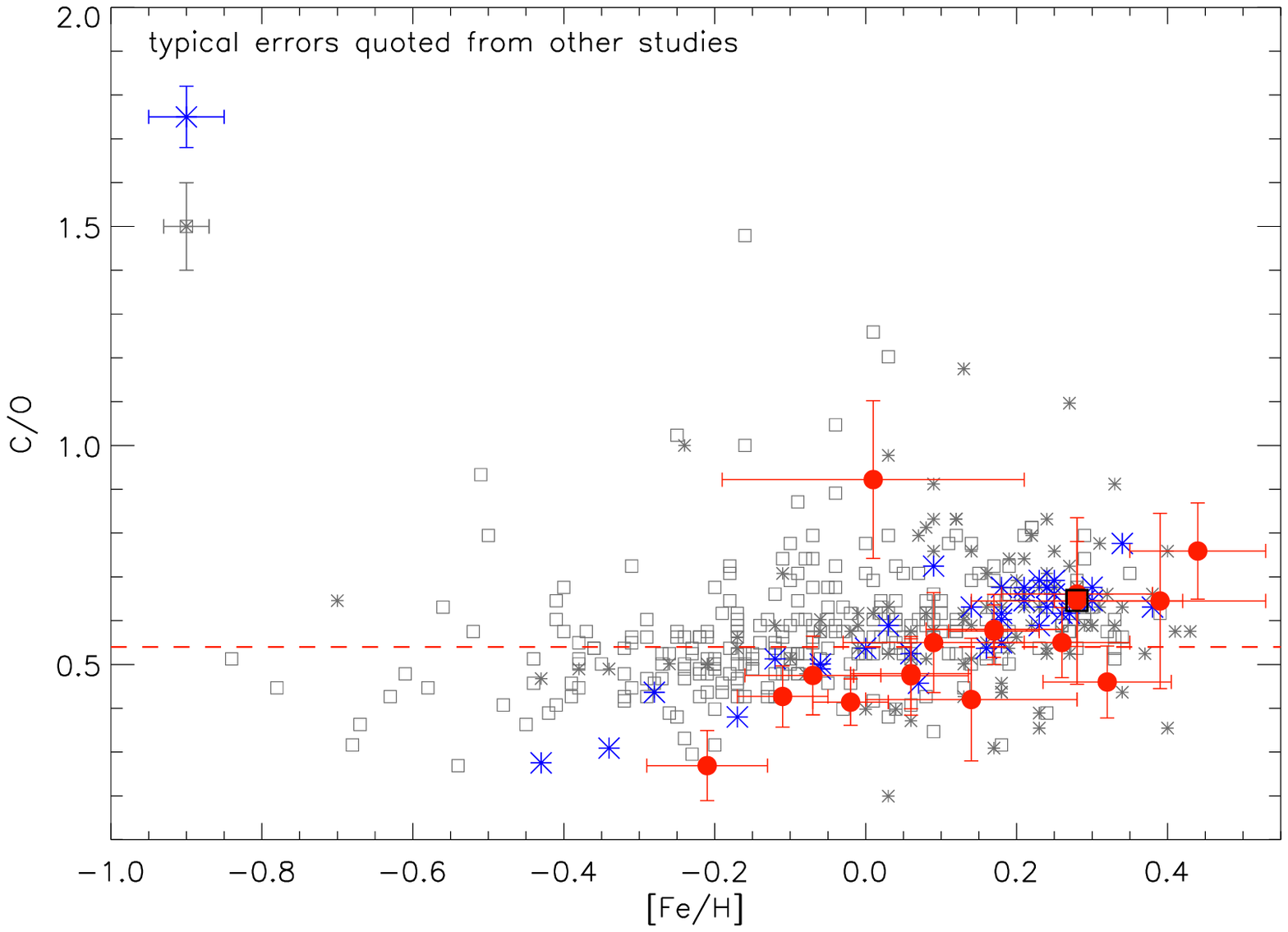}}
\caption{C/O versus [Fe/H] from Delgado Mena et al.\,(2010) and Nissen\,(2013) [all
  Nissen\,(2013) hosts are in the Delgado Mena et al.\,(2010) host
  sample]. Colors and symbols are the same as in Figure
  \ref{comparison_plot1}. Left: C/O ratios as reported in respective
  sources, using their C/O$_{\odot}$ (see text for discussion). Dashed
  lines show the C/O$_{\odot}$ adopted by each
  source. Right: All C/O ratios normalized to the same C/O$_{\odot}$
  adopted in this work, C/O$_{\odot}$=0.54 (log$N$(C)$_{\odot}$=8.39,
  log$N$(O)$_{\odot}$=8.66; Asplund et al.\,2005).}
\label{comparison_plot2}
\end{figure}

Figure \ref{comparison_plot2} illustrates values of C/O versus [Fe/H]
from this study, along with those values from Delgado Mena et
al.\,(2010) and Nissen (2013). All three studies find a clear increase
in C/O verus [Fe/H], which represents the signature of Galactic
chemical evolution, as discussed previously. The relation in C/O in
this work falls somewhat below those of the other two studies, but all
three exhibit similar slopes. This similarity is born out by a
quantitative comparison of C/O versus [Fe/H] between Nissen (2013) and
this study. Nissen derived a linear fit of C/O = 0.54[Fe/H] + 0.56, while the
same fit to the results here find C/O = 0.53[Fe/H] + 0.45.  
The adopted
solar value in this study of C/O = 0.54 is larger by 0.09  
than the value
defined by the best-fit linear relation defined by our sample of stars and
our results.  Including the outlier HD 189733 in our
  linear fit results in C/O = 0.43[Fe/H] + 0.49, corresponding to a
  C/O ratio 0.05 smaller than the solar value of 0.54; including this
  outlier increases the scatter around the fit from 0.04 to 0.1. A linear difference of 0.09  
corresponds to 0.08  
dex for the solar-relative [C/O], which is comparable to the repsective offsets of -0.08 dex and +0.01 dex in the [C/H] and [O/H] relations.    

Another way of investigating the inherent scatter
within our results is to remove the linear best-fit from the values of
C/O and look at the scatter about the fitted relation. When this is
done the median residual scatter in C/O is $\pm$0.04
, or 0.03 dex in [C/O].
This comparison of C/O versus [Fe/H] trends between Nissen (2013) and this
study indicates that the derived slopes are very similar, but there remain
small offsets in zero-point C/O of $\sim$ 0.10 - 0.15 caused by a combination
of differences in (presumably) the stellar samples, the adopted solar
C/O ratios (0.58 for Nissen and 0.54 for this study), as well as the
abundance analysis, e.g., much of the offset is due to somewhat smaller
values of [C/H] at our lower metallicity range. 

The mean and standard deviation of the [C/H],
[O/H], and C/O distributions from this study, as well as those from
all the previous studies of host star carbon and oxygen abundances
mentioned above, are listed in Table \ref{tab:compare2}. 
The mean [C/H] of the transiting exoplanet hosts in this
paper is  
less than the mean [C/H]$_{\rm{host}}$ from the
previous works, 0.14 in the five previous studies versus 0.08 found here. 
The mean [O/H] value found in our sample is the same as 
the mean
[O/H]$_{\rm{host}}$ from previous studies, 
0.07. 
However, the standard
deviations of [C/H]$_{\rm{transiting}}$ and [O/H]$_{\rm{transiting}}$
from this paper are large, 0.20 and 0.13, respectively, 
so any differences in our mean [C/H] and [O/H] values are to be viewed
with caution. 
The mean C/O ratio of the transiting
exoplant host stars in our sample is  
0.54, with a standard deviation of 0.15, versus the mean from the
previous papers of 0.71 with a standard deviation of 0.07. 
Therefore, the sample of carbon and oxygen abundance ratios for transiting exoplanet
host stars presented here, while marginally consistent, are on average
lower than those  
measured by other groups for non-transiting exoplanet host
stars. 
Our measurements are more in line with the suggestions by
Fortney\,(2012) and Nissen (2013) that prior studies overestimated C/O
ratios; the mean C/O$_{\rm{hosts}}$ of Nissen is 0.63$\pm$0.12.  

However, as noted by Fortney\,(2012), each previous study scales their
C/O ratios based on different log$N$(C)$_{\odot}$ and
log$N$(O)$_{\odot}$ values. Delgado Mena
et al.\,(2010) list log$N$(C)$_{\odot}$ and
log$N$(O)$_{\odot}$ as 8.56 and 8.74, respectively, resulting in
C/O$_{\odot}$=0.66. These are also the values listed in Ecuvillion et
al.\,(2004) and (2006), the quoted
sources of Bond et al.\,(2010). Petigura \& Marcy\,(2011) list log$N$(C)$_{\odot}$ and
log$N$(O)$_{\odot}$ as 8.50 and 8.70, respectively, resulting in
C/O$_{\odot}$=0.63. Nissen\,(2013)'s
C/O$_{\odot}$=10$^{8.43}$/10$^{8.665}$=0.58. Figure
\ref{comparison_plot2} illustrates the different C/O$_{\odot}$ from Delgado Mena
et al.\,(2010) and Nissen\,(2013). Accounting for the
difference in log$N$(C)$_{\odot}$ and
log$N$(O)$_{\odot}$ decreases the average C/O ratios from the other sources
(from top to bottom) in
Table \ref{tab:compare2} by $\sim$0.15, $\sim$0.13, $\sim$0.11, and
$\sim$0.05, closer to the average C/O ratio we derive for our
sample. Figure \ref{comparison_plot2}'s
right panel shows the C/O ratios of Delgado Mena et al.\,(2010) and
Nissen\,(2013) along with those derived in this work, all on the same
scale, illustrating how using different solar C and O
absolute abundances changes the resulting C/O ratios. This underscores the
caution, as mentioned in Fortney\,(2012) and Nissen\,(2013), required
when directly comparing C/O ratios derived from different groups. 

We now focus on the C/O ratios in each studied system to investigate possible
links between host star C/O ratios with planetary and system properties. 

\subsection{Trends with C/O$_{\rm{host\,star}}$ versus Planetary Parameters}

Presently there are two major observed trends relating stellar chemical
composition to the presence of planets -- hot Jupiter exoplanets are
more often found around intrinsically higher-metallicity stars (e.g.,
Fischer \& Valenti 2005), and the fraction of stars with giant planets
increases with stellar mass (e.g., Johnson et al.\,2010; Ghezzi et
al.\,2010;  
Gaidos et al.\,2013). Measuring potential host stars' chemical
abundance distributions may develop into a powerful tool for
inferring the presence, or even specific type (size, orbit, composition),
of exoplanets around different types of stars. This technique is of
increasing importance in the context of large surveys that are
discovering exoplanets, and targeted studies of unusual or potentially-habitable exoplanets. 

\begin{figure}[hbt]
\figurenum{5}

   \subfigure{\includegraphics[width=0.5\textwidth]{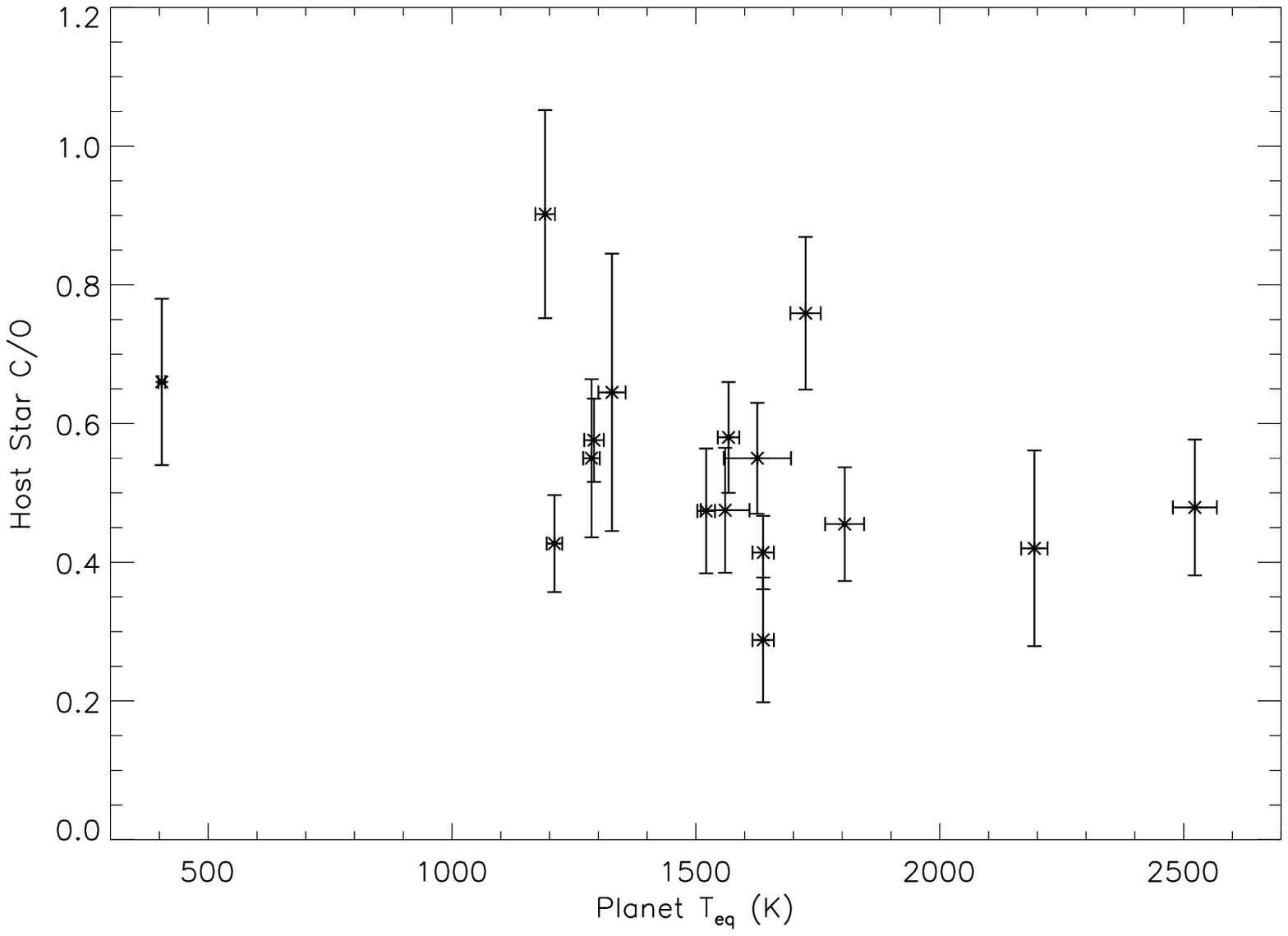}}
\quad
   \subfigure{\includegraphics[width=0.5\textwidth]{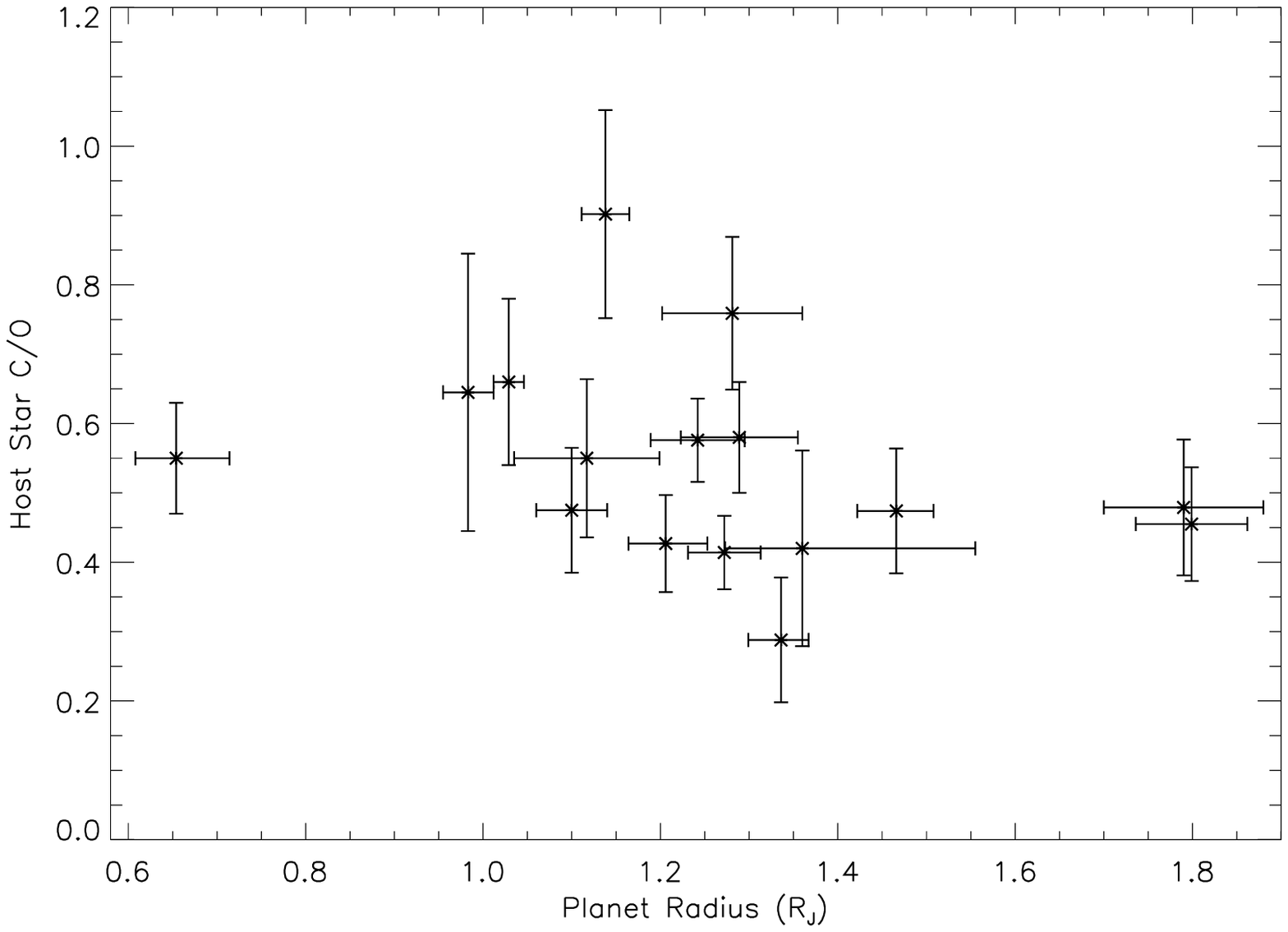}}
\caption{Host star C/O ratio versus planetary equilibrium temperature (left) and radius (right). The planetary parameters are from
  the NASA Exoplanet Archive, and the host star C/O ratios are
  derived in this paper.}
\label{pparam}
\end{figure}

In this study we explore whether the stellar C/O ratio has
predictive power with respect to hot Jupiter properties, particularly
the exoplanetary atmosphere compositions.  
Characteristic observations
of the atmospheres of the exoplanets in this sample -- the
\textit{Spitzer/IRAC} 3.6, 4.5, 5., and 8.0 $\mu$m secondary eclipse fluxes -- as well as their physical
properties like mass, radius, semi-major orbital axis, period, and
equlibrium temperature were gathered from the NASA Exoplanet
Archive\footnote{http://exoplanetarchive.ipac.caltech.edu/} and
compared to host star C/O ratios. By eye it appears that
planet radius and planet equlibrium temperature may decrease with
increasing C/O$_{\rm{host\,star}}$ (Fig.\ref{pparam}), but these trends are dominated by one
or two 
points and, once these points are removed, no
significant trends with planet parameters are found. We also find weak
  negative correlations between each system's C/O$_{\rm{host star}}$ and planetary
  \textit{Spitzer}/IRAC secondary eclipse fluxes (e.g., $r\sim$-0.4 to
  -0.6), but these correlations are not statistically significant ($p>$0.05). 
  
This lack of trends between C/O$_{\rm{host\,star}}$ is perhaps not
surprising. The hot Jupiter host stars in this sample were chosen
based on the amount of observational data that exists for their
planets, and thus how ``characterizable'' their planets' atmospheres
are, with the goal of directly comparing star and planet
C/O ratios. No planetary or stellar parameters serve as ``control
variables'' in this study, and our sample is actually diverse in both respects. The
host stars span 5100$\lesssim$T$_{\rm{eff}}\lesssim$6470 K,
-0.21$\lesssim$[Fe/H]$\lesssim$0.44, and spectral types F6 through
K1. The planets in our sample range in mass from $\sim$0.5-4.2 M$_{\rm{J}}$, in period
from $\sim$1.09-111 days (with the second longest being 4.5 days),
in density from $\sim$0.2-8 g cm$^{-3}$ (with the second most-dense
being 3.4 g cm$^{-3}$), and in equilibrium temperature from
$\sim$400-2500 K. That we do not find a significant correlation between
C/O$_{\rm{host\,star}}$ and any of these planetary parameters implies
that (1) our sample may yet be too small to reveal distinct trends, and/or (2) the influence of the host star C/O ratio is a more complex
function of multiple parameters of
the planet and/or its formation history.

While (1) is possible, (2) also seems likely and could result
in the C/O ratio comparison between stars and planets serving a more
interesting function. In protoplanetary disks, different
condensation fronts due to temperature and the movement of gas and
grains in the disk can change the relative ratios and/or
of carbon and oxygen 
as compared to those in the parent star (Stevenson
\& Lunine 1988; Lodders 2010; Ciesla \& Cuzzi 2006; \"Oberg et al. 2011). In particular, the enhancement or depletion of water and thus oxygen is sensitive
to the size and migration of icy solids in the disk, so the C/O ratios of
the inner and outer disk regions evolve with time and depend on
both initial conditions and the efficiency with which solids grow to
large sizes (Ciesla \& Cuzzi 2006; Najita et al.\,2013). Overall, the final
C/O$_{\rm{planet}}$ does not necessarily reflect the
C/O$_{\rm{disk-average}}$, and depends 
on the location and timescale of formation, how much of the atmosphere is accreted from
gas versus solids, and how isolated the atmosphere is from mixing with
core materials
(Ciesla \& Cuzzi 2006; \"Oberg et al.\,2011). In our own solar system
gas giant planets, oxygen is not well constrained because water, the
major oxygen carrier, condenses deeper down in their cool (T$\leq$125
K) atmospheres, out of the observable range of remote spectra
(Madhusudhan 2012). However, carbon is known to be 
enhanced above solar by factors of $\sim$2-6, 6-11, 18-50, and 28-63 in
Jupiter, Saturn, Uranus, and Neptune, respectively (Wong et al.\,2008
and references therein). Thus, though
the composition of the host star provides a good estimate of the
system C/O ratio and the natal molecular cloud environment,
differences between the host star and planetary C/O ratios may be
common, and may be used to probe where and when in the disk the planet formed.

A third possibility is that the host star C/O ratio has no connection
to the formation of planets and is not a useful metric for
distinguishing planet types. However, theoretical
results (e.g., Johnson et al.\,2012; Ali-Dib et al.\,2013) demonstrating the influence of the host star C/O ratio on
the composition of protoplanetary disk, 
 and recent observations (e.g., Najita et al.\,2013; Favre et
 al.\,2013) indicating that disks themselves likely have a range of C/O
 ratios which are related to other planet formation parameters
 (mass of the disk, grain growth and composition, etc.), suggest that C/O ratios of host stars do play a role, at some stage, in planet
 formation.

\subsection{Carbon and Oxygen in Specific Exoplanet Systems}
A small fraction of exoplanets, mostly hot
Jupiters orbiting very close to their host stars, have been
observed and analyzed with spectroscopy and photometry in the optical
and near-infrared during primary transit (e.g., Madhusudhan \& Seager
2009; Swain et al.\,2008, 2013;  Moses et al.\,2011; Mandell et
al.\,2013) 
and/or secondary eclipse  (e.g., Charbonneau et al.\,2005, 2008;
Knutson et al. 2008; Madhusudhan et al.\,2011; Crossfield et al.\,2012). 
Direct imaging of exoplanets in wider orbits 
(e.g., Marois et al.\,2008 \& 2010; Lagrange et al.\,2009; Bailey et al.\,2014) has also opened up for study a new population of self-luminous planets in Jovian-type orbits.  

As discussed in the introduction, a gas giant planet's C/O ratio has
important implications for its composition. At the temperatures and pressures
characteristic of such atmospheres, a high C/O ratio
($\gtrsim$0.8) can
significantly alter the temperature and chemistry structure by
depleting the dominant opacity source H$_2$O and introducing new
sources that are C-rich like CH$_4$, HCN, and/or other hydrocarbons.  
In thermochemical equilibrium, C/O$>$1 causes O to be confined
mostly to CO, depleting H$_2$O and enhancing CH$_4$ versus what is
expected in solar-abundance atmospheres (C/O$_{\odot}$=0.55$\pm$0.10; Asplund et
al.\,2009; Caffau et al.\,2011), which have abundant H$_2$O and CO 
(Madhusudhan 2012; Moses et al.\,2013). In carbon-rich atmospheres, the temperature controls how
depleted the H$_2$O is compared to solar and the partitioning of
carbon between CH$_4$ and CO, which in turn influences the oxygen
balance between CO and H$_2$O 
(Madhusudhan 2012).

With the ability to constrain exoplanet atmosphere
compositions (e.g., 
Madhusudhan 2012; Lee et al.\,2012; Moses
et al.\,2013; Konopacky et al.\,2013; Line et al.\,2013), a logical next step towards 
determining the host star's influence on exoplanet formation is the
direct comparison of the abundance ratios of star/planet
pairs.

\subsubsection{WASP-12}
For WASP-12b, one of the brightest transiting exoplanets, the comparison between host star and planet composition has
already begun (Madhusudhan et al.\,2011;
Madhusudhan 2012; Petigura \& Marcy 2011; Crossfield et al.\,2012;
Swain et al.\,2013; Copperwheat et al.\,2013; 
Sing et al.\,2013). The host star is found in this work to have [Fe/H] =
0.06$\pm$0.08 and C/O=0.48$\pm$0.08. We note that this metallicity
differs significantly from the [M/H]= 0.30$^{+0.05}_{-0.10}$ reported by Hebb et
al.\,(2009) in the WASP-12b discovery paper,
based on spectral synthesis of four regions including the Mg b
triplet at 5160-5190\,{\AA}, Na I D doublet at 5850-5950\,{\AA}, 6000-6210
\AA, and H$\alpha$ at 6520-6600\,{\AA}, following the procedure of
Valenti \& Fischer\,(2005). 

Coupling their
atmospheric modeling and retreival methods to published secondary eclipse
photometry and spectroscopy spanning 0.9 $\mu$m to 8 $\mu$m, 
Madhusudhan et al.\,(2011)
and Madhusudhan\,(2012) suggest that WASP-12b's atmosphere has a C/O
ratio$\geq$1. Their best fit describes an atmosphere abundant in CO, depleted in H$_2$O, and enhanced
in CH$_4$, each by greater than two orders of magnitude compared to
the authors' solar-abundance, chemical-equilibrium models. However, this
 high 
C/O$_{\rm{planet}}$ ratio for WASP-12b's atmosphere is ruled out at the $>$3$\sigma$ level with new
observations at 2.315 $\mu$m and reanalysis of previous observations
accounting for the recently
detected close M-dwarf stellar companion (Bergfors et al.\,2011; Crossfield et al.\,2012). Including the dilution of the reported
transit and eclipse depths due to the M-dwarf, the dayside spectrum of
WASP-12b is best explained by a featureless 3000 K blackbody (Crossfield et
al.\,2012). Subsequent data (Sing et al.\,2013) do not detect metal
hydrides MgH, CrH, and TiH or any Ti-bearing molecules, which were
previously suggested as indicative of high-C/O ratio scenarios
(Madhusudhan 2012; Swain et al.\,2013). 

A C/O$<$1 composition for WASP-12b is also consistent with the study of Line et al.\,(2013), who use a
systematic temperature and abundance retrieval
analysis, combining differential evolution MCMC
with an optimal-estimation-based prior, to rule
out strong temperature inversion in WASP-12b's atmosphere and
thus the presence of TiO causing such an inversion. Accounting for the
M dwarf companion, these authors determine a best-fit C/O ratio for WASP-12b of 0.59
($\chi^{2}_{best}$/$N$=2.45, with a 68\% confidence
interval of 0.54-0.95), suggesting that a high C/O ratio is not
the explanation for WASP-12b's lack of atmospheric temperature
inversion. If WASP-12b's C/O ratio really is super-solar and
significantly different than its host star (0.48$\pm$0.08), this
suggests that some other mechanism influenced the composition of the
exoplanet during its formation/evolution. \"Oberg et al.\,(2011)
note that the high C/O and substellar C/H reported by
Madhusudhan et al.\,(2011) are only consistent with an atmosphere
formed predominantly from gas accretion outside the water
snowline. With our updated metallicity measurement, C/H in WASP-12
decreases to $\sim$ 5$\times 10^{-4}$, exactly in the middle of the C/H
distribution spanned for the planet in Madhusudhan et al.\,(2011)'s
best-fitting ($\chi^2<$7) models. Thus, by these models, the planet's
C/H is just as likely to be substellar as super-stellar. 
More data, particularly around 3 $\mu$m (see Line et al.\,2013, Figure
1), can help further constrain WASP-12b's C/O ratio and enable a more
meaningful comparison between planet and host star. We note that the very
recent HST/WFC3 transit spectra of WASP-12b from 1.1-1.7 $\mu$m reported by Mandell et
al.\,(2013) are fit equally well by oxygen- and carbon-rich models of
Madhusudhan et al.\,(2012).

\subsubsection{XO-1}
The hot Jupiter XO-1b's (McCullough et al.\,2006)
four
\textit{Spitzer/IRAC} photometric secondary eclipse observations have been explained with a
solar-composition, thermally-inverted model (Machalek et al.\,2008). However,
it is also possible to fit the observations with a non-inverted
(Tinetti et al.\,2010), 
potentially carbon-rich 
atmosphere model (Madhusudhan
2012), which may include disequilibrium chemistry like photochemistry and/or
transport-induced quenching (Moses et al.\,2013). 
As the favored C/O$\geq$1 
models 
are heavily dependent on the 5.8 $\mu$m photometric point, new observations are necessary to confirm the
carbon-rich nature of XO-1b's atmosphere. Here we find in the XO-1
host star [Fe/H] = -0.11$\pm$0.06, with [C/H] =
-0.19$\pm$0.04 and [O/H] = -0.09$\pm$0.05, resulting in C/O =
0.43$\pm$0.07. 

\subsubsection{TrES-2 and TrES-3}
TrES-2b and TrES-3b were among the first transiting hot Jupiter exoplanets discovered 
(O'Donovan et al.\,2006; O'Donovan et al.\,2007). 
Both fall under the highly-irradiation
``pM'' class predicted to have temperature inversions in their upper
atmospheres (Fortney et al.\,2008). Secondary eclipses of TrES-2b and
TrES-3b were observed with the CFHT Wide-field Infrared
Camera 2.15 $\mu$m filter, (Croll et al.\,2010a; 2010b), in
\textit{Spitzer}/IRAC's four near-IR bands (O'Donovan et al.\,2010;
Fressin et al.\,2010, respectively). Radiative transfer
analyese of Line et al.\,(2014) use the CFHT and $\textit{Spitzer}$
data indicate a range of temperature-pressure profiles are 
too cool for TiO and VO to be in the gas phase, which suggests 
that these species do not cause thermal inversions.  
Line et al.\,(2014) find that the data provide minimal
constraints on the abundances of H$_2$O, CO$_2$, CO, and CH$_4$, and
thus TrES-2b's atmospheric C/O ratio. 
Their best fit ($\chi^{2}_{best}$/$N$=0.60) is
0.20, but their 68\% confidence interval spans
0.021-8.25. Interestingly, the C/O 
ratio of TrES-2 that we derive, 0.41$\pm$0.05, has the lowest error in our sample and also the second-lowest C/O ratio value in our
sample. Hence, if TrES-2b accreted much of its gas from a reservoir
similar in composition to its host star, and its atmosphere remained
mostly isolated from its interior, it may also have a sub-solar
atmospheric C/O ratio.

For TrES-3b, radiative transfer analyses of the
infrared photometry indicate
that H$_2$O is well determined with an
abundance near 10$^{-4}$, and CH$_4$ has an upper limit
of $\sim$10$^{-6}$ (Line et al.\,2014). CO$_2$ shows a weak upper limit $\sim$10$^{-4}$, derived from the 2.1 $\mu$m CO$_2$ band wings within the $K$ band
measurement, while CO is unconstrained due to the large uncertainty in
the $Spitzer$ 4.5 $\mu$m data point, combined with the fact that no other molecular
absorption features of CO are probed by the current data (Line et
al.\,2014). Line et al.\,(2014) infer that C/O$>$1 in TrES-3b due to
the relatively high-confidence limit on H$_2$O and the small upper
limit on CH$_4$. However, the data provide no constraints on the CO
abundance, which is expected to be the major carbon carrier in an
atmosphere as hot as TrES-3b. Their best fit
($\chi^{2}_{best}$/$N$=0.067) C/O ratio for TrES-3b is 0.22, with a
68\% confidence interval of 0-0.97. The very recently published HST/WFC3 secondary
eclipse observations of TrES-3b are poorly fit with a
solar-composition model ($\chi^{2}/N=3.04$), whereas the WFC3 data
plus the existing $\textit{Spitzer}$ photometry are more consistent
($\chi^{2}/N=0.75$) with an atmosphere model depleted in CO$_2$ and
H$_2$O by a factor of 10 relative to a solar-composition model (Ranjan et al.\,2014). 

TrES-3's C/O ratio
derived here, 0.29$\pm$0.09, also has a small error and is the
lowest C/O ratio in the hot Jupiter host stars studied here. The
large span in the planet's C/O ratio found by Line et al.\,(2014) is still too large to draw meaningful conclusions about the
formation location and/or growth history of TrES-3b. However, if the
degeneracy between CO and CO$_2$ absorption in the \textit{Spitzer}
4.5 $\mu$m data point is broken by, for instance, observations of the
2.6 $\mu$m or 15 $\mu$m CO$_2$ bending band or the 5 $\mu$m CO
fundamental band by SOFIA/FLITECAM (McLean et al.\,2006) or SOFIA/FORCAST (Adams et
al.\,2010), both the CO and CO$_2$ contributions could be
better estimated and lead to a tighter C/O ratio constraint for
TrES-3b. This system is intriguing and important for further
investigation due to both our firmly sub-solar C/O ratio and the
relatively metal-poor nature of the host star ([Fe/H]=-0.21$\pm$0.08,
the lowest in our sample), which distinguishes TrES-3 from most other hot Jupiter hosts. 
  
\subsubsection{HD 149026}
Observational constraints and extensive
theoretical modeling indicate that the exoplanet HD 149026b has between 45-110
M$_{\oplus}$ of heavy elements 
in its core and
surrounding envelope (Sato et al.\,2005; Fortney et al.\,2006; Ikoma
et al.\,2006; 
Broeg \& Wuchterl 2007)
, making the core of HD 149026b at least twice as massive
as Saturn's,  
even though its radius is
$\sim$0.86 R$_{\rm{Saturn}}$ (Triaud et al.\,2010) and its mass is
$\sim$1.2 M$_{\rm{Saturn}}$ (Sato et al.\,2005). The massive core of
HD 149026b challenges formation by traditional core accretion theory, and 
many modified formation scenarios have been suggested, 
including collision with an outer additional giant planet (Sato et
al.\,2005; Ikoma et al.\,2006), accretion of planetesimals or smaller
(super-Earth-sized) planets (Ikoma et al.\,2006; Broeg \& Wuchterl
2007; Anderson \& Adams 2012), or  
core accretion in
a disk with $\times$2 the heavy element mass in the solar nebula
(Dodson-Robinson \& Bodenheimer 2009). This latter explanation stems
from the metal-rich nature of the star -- more
massive/metal-rich disks form planets more readily (Ida \& Lin 2004ab) and metal-rich planets tend to be associated with metal-rich
stars (Guillot et al.\,2006; Burrows et al.\,2007a; Miller \& Fortney
2011). Here we find [Fe/H]=0.26$\pm$0.09 for HD 149026, which is not
as high as previous studies ([Fe/H]=0.36$\pm$0.05; Sato et
al.\,2005), but still suggests that, overall, the initial metal
abundance in the molecular cloud/disk was enhanced above
solar. We measure [C/H]$=$0.26$\pm$0.08 and
[O/H]$=$0.25$\pm$0.04, both enhanced above solar, resulting in a C/O
ratio of 0.55$\pm$0.08, consistent with solar. 

Stevenson et al.\,(2012)
find that the \textit{Spitzer} secondary eclipse observations of HD
149026b (at 3.6, 4.5, 5.8, 8.0, and 16 $\mu$m) can be fit using 
models with an atmosphere in chemical equilibrium and lacking a temperature
inversion, with large amounts of CO and CO$_2$, and a metallicity
$\times$30 solar (Fortney et al.\,2006). 
The retrieval
results of Line et al.\,(2013) also 
indicate the atmosphere of HD 149026b has more CO and CO$_{2}$ than
CH$_{4}$, which makes sense given the planet's high temperature
($\sim$1700 K) that favors formation of CO over CH$_{4}$ at solar
abundances. There is also a peak in the Line et al. modeled composition probability
distribution of the H$_2$O mixing ratio near $\sim$10$^{-5}$. Given this H$_2$O
abundance, and the low abundance of CH$_{4}$, the C/O ratio of HD
149026b is likely $<$1, but is remains poorly constrained 
(0.55, with a $\chi^{2}_{best}$/$N$=0.23 fit and a 68\% confidence
interval of 0.45-1.0; Line et al.\,2013). A better estimate of HD 149026b's
C/O ratio as compared to the C/O ratio of its host star
(0.55$\pm$0.08) may shed light on the planet's history and the origin
of its massive core. 
This heavy-cored hot Jupiter system, with the host star carbon and
oxygen abundances presented here, is 
a valuable test-bed for studying how massive planets form.

\subsubsection{XO-2} 
The hot Jupiter XO-2b has a host star, XO-2N, with a binary companion,
XO-2S, located $\sim$4600 AU away and not known to host a hot
Jupiter-type planet (Burke et al.\,2007). The stars are of
similar stellar type, meaning that the non-hosting companion
can be used to check for effects of planet formation on the host star,
e.g., stellar atmospheric pollution. 
XO-2b 
has been observed with
with HST and \textit{Spitzer} (Machalek et al.\,2009; Crouzet et
al.\,2012) as well as from the ground (Sing et al.\,2012 \& 2011;
Griffith et al.\,2014). Griffith et al.\,(2014) find, with a
comprehensive analysis of all existing data, that the water abundance
that best matches most of the data is consistent with an atmosphere
that has the same metallicity and C/O ratio as the host star in
photochemical equilibrium. However there are outlying observations, so
additional measurements and needed to understand the cause for the
outliers and to investigate the carbon abundance in XO-2b.

Teske et al.\,(2013b) derived the carbon and oxygen
abundances of both binary components, and found [C/H]$=$
$+$0.26$\pm$0.11 in XO-2S versus $+$0.42$\pm$0.12 in XO-2N, and [O/H]$=$
$+$0.18$\pm$0.15 in XO-2S versus $+$0.34$\pm$0.16 in XO-2N. The
stars are enhanced above solar in C and O, with XO-2N being slighly
more carbon- and oxygen-rich. Their 
relative enhancements result in both having C/O=0.65$\pm$0.20. (Note
that this value is slightly larger than that reported in Teske et
al.\,2013b because the log$N$(O)$_{\odot}$ in this work is 8.66 versus
8.69 in Teske et
al.\,2013b.) Both XO-2N and XO-2S fall exactly on our linear trends
with [Fe/H] discussed in $\S$4.1 ([Fe/H]$_{XO-2N}$=0.39$\pm$0.14,
[Fe/H]$_{XO-2N}$=0.28$\pm$0.14). The elevated-above-solar [C/H] and [O/H] values in
the two stars are strong evidence that their parent molecular
cloud was elevated in both carbon and oxygen. Given that their C/O
ratios are identical, the key to understanding why XO-2N has a planet
and XO-2S does not may lie in the exoplanet
composition. 

\subsubsection{CoRoT-2}
Of the planets around the host stars in our sample, CoRoT-2b is
perhaps the most puzzling in terms of its atmospheric
structure. Traditional solar composition, equilibrium
chemistry models are unable to reproduce the unusual flux ratios 
from the three \textit{Spitzer} channel observations (it is missing
5.8 $\mu$m) of this very massive hot
Jupiter (Alonso et al.\,2010; Gillon et al.\,2010; Deming et al.\,2011;
Guillot \& Havel 2011). Despite its large mass, the planet has one of the greatest radius
anomalies -- slower-contraction evolution models that explain the
radius anomalies of other inflated planets cannot justify this case (Guillot \& Havel 2011). Furthermore, the host star
is young (formed within 30-40 million years; Guillot \& Havel 2011),
chromospherically active and the system has been suggested to be undergoing magnetic star-planet interactions due to the observed
stellar spot oscillation period that is $\sim$10$\times$ the synodic
period of the planet as seen by the rotating active longitudes (Lanza
et al.\,2009).

CoRoT-2b's emission data are difficult to interpret, largely
  because of the anomalously high 4.5/8.0 $\mu$m flux
  ratio. Excess CO mass loss has been suggested to enhance the 4.5
  $\mu$m flux, as has some unknown absorber acting only below $\sim$ 5
  $\mu$m (Deming et al.\,2011; Guillot \& Havel\,2011). Alternatively
  the low 8 $\mu$m flux may be caused by a high C/O ratio through
  absorption of CH$_4$, HCN, and C$_2$H$_2$
absorption (Madhusudhan\,2012). In addition, the lack of a 5.8 $\mu$m
measurement leads to a poor constraint on the H$_2$O abundance, which
strongly dictates the resulting C/O ratio. Wilkins et
al.\,(2014) find that no single atmospheric model is able to reproduce
of all the available CoRoT-2b data, including their new 1.1.7 $\mu$m HST/WFC3
spectra, the optical eclipse observed by
\textit{CoRoT} (Alonson et al.\,2009; Snellen et al.\,2010) and the
previously-modeled infrared photometry. More complex
models with differing C/O ratios or varying opacity sources do not provide a
fit more convincing that a one-component blackbody, which in itself
still misses the ground-\textit{Spitzer} eclipse amplitudes by
$\sim$1.8$\sigma$ (Wilkins et al.\,2014).

Disequilibrium chemistry can significantly affect CoRoT-2b's atmospheric composition. For
instance, for a high C/O ratio,
H$_2$O is predicted to be enhanced above 10$^{-2}$ bar by $\sim$four
orders of magnitude due to both transport-induced quenching and CO
photochemistry in the upper atmosphere (Moses et al.\,2013). HCN
and other C$_x$H$_x$ compounds may also
result from the reaction of the leftover C with N or H$_2$ (Moses
et al.\,2013). Disequilibrium chemistry models with 0.5$\times$solar
metallicity, moderate mixing, 
 and C/O=1.1 yield a
significantly better match to the four CoRoT-2b infrared secondary
eclipse observations, providing a $\chi^2/N$=1.3, versus the
solar-composition models, which provide a $\chi^2/N$=7.2 (Moses et
al.\,2013).

The host star C/O ratio derived here,
0.47$\pm$0.09, is equal to C/O$_{\odot}$ within error, as is the
overall metallicity of the host star, [Fe/H]=0.06$\pm$0.08. We note that the [C/H] value
measured here for CoRoT-2 is based on only 2 carbon lines (5380\,{\AA} and
7113\,{\AA}), and the [O/H] value is based on only the O I triplet at
$\sim$7775\,{\AA}; the other potential C and O lines were too weak to
be reliably measured in our data. In addition, while the O I triplet
NLTE corrections are nominally valid in this case because the
T$_{\rm{eff}}$ of CoRoT-2 is $\geq$5400, implementing these
corrections results in an [O/H] that is larger than the LTE case, the
opposite direction of the corrections at near solar temperatures. 
However,[O/H]$_{\rm{LTE}}$=0.02, the same as [O/H]$_{\rm{NLTE}}$=0.06
 within error (0.07 dex); if we adopt the [O/H]$_{\rm{LTE}}$ value,
 CoRoT-2's C/O ratio increases by only 0.05 dex, also within error. 

CoRoT-2b's C/O ratio, while still uncertain, could
plausibly be $>$1. Several different scenarios could account for
C/O$_{\rm{planet}}>$C/O$_{\rm{star}}$. CoRoT-2b could have accreted carbon-rich very hot gas from
the inner disk regions ($\lesssim$0.1 AU). Alternatively, the planet could have accreted
the majority of its gas from beyond the H$_2$O snow line (causing it
to be oxygen-depleted), or accreted solid material depleted in oxygen (e.g., from a ``tar line'' inward
of the snow line; Lodders 2004). Interestingly, to explain CoRoT-2b's
inflated size and large mass, Guillot \&
Havel\,(2010) propose that the CoRoT-2 system previously
included multiple giant planets that collided within the last
$\sim$20 million years to create the currently-observed CoRoT-2b. This scenario could result in a planet that differs
 significantly in composition from the original states of the impactors, potentially erasing the signatures of where/from what
 material in the disk the planet formed. CoRoT-2b is another candidate
 for which additional SOFIA observations at near- (2.6 $\mu$m, 6
 $\mu$m) and mid-infrared ($>$20 $\mu$m) wavelengths can help better
 constrain the exoplanet C/O ratio and thus its formation history.      

\subsubsection{HD 189733}
The hot Jupiter HD 189733b is one of the best-studied to date, with data
spanning $\sim$0.3-24 $\mu$m (Barnes et al.\,2007; Grillmair et al.\,2007; 2008; Tinetti et
al.\,2007; Knutson et al.\,2007; 2009; 2012; Redfield et al.\,2008; Charbonneau et
al.\,2008; Beaulieu et al.\,2008; Pont et al.\,2008; D{\'e}sert et
al.\,2009; Swain et al.\,2008 \& 2009a; Sing
et al.\,2009 \& 2011; Agol et al.\,2010; Gibson et al.\,2012; Evans et al.\,2013; Birkby et al.\,2013). The first complete
atmospheric study via statistical analysis with a systematic, wide parameter
grid search (Madhusudhan \& Seager 2009) analyzed separately
spectroscopic data from 5-14 $\mu$m (Grillmair et al.\,2008),
photometric data at 3.6, 4.5, 5.8, 8, 16, and 24 $\mu$m (Charbonneau
et al.\,2008), and spectrophotometric data from 1.65-2.4 $\mu$m (Swain
et al.\,2009a). Madhusudhan \& Seager\,(2009) place constraints at the
$\xi^2$=2 level (where $\xi^2$ is a proxy for the reduced $\chi^2$
using the \# of data points as $N$) on HD
189733b's atmospheric mixing ratios of H$_2$O, CH$_4$, and CO$_2$
using the spectrophotometric data, as it includes features of all
of these molecules as well as CO. Their resulting C/O ratio range for HD
189733b is between 0.5 and 1. 

Subsequent analysis of all of the available infrared secondary eclipse
measurements with the Bayesian optimal estimation retrieval scheme
NEMESIS (Irwin et al.\,2008) placed constraints on molecular
abundances ratios of HD 189733b's atmosphere, resulting in a
best-estimate C/O ratio between 0.45 and 1 for $\xi^2<$0.5 and between
0.15 and 1 for $\xi^2<$2 (Lee et al.\,2012). However, these authors
caution that the current secondary eclipse data are only able to
constrain the thermal structure of HD 189733b at some pressure levels,
and the mixing ratios of H$_2$O and CO$_2$ with large uncertainties
ranging between 9-500$\times$10$^{-5}$ and 3-150$\times$10$^{-5}$
for $\xi^2<$0.5, respectively,
due to the model degeneracies. The most significant degeneracy they
find is between temperature and H$_2$O abundance at 300 mbar
pressure. 

The H$_2$O abundance has the
biggest influence on the overall shape of a hot Jupiter spectrum in
thermochemical equilibrium. Moses et al.\,(2013)
focuses on the H$_2$O mixing ratio constraint of
$\sim$1$\times$10$^{-4}$ from Madhusudhan \& Seager\,(2009)
in their exploration of disequilibrium chemistry 
using
the combined data sets mentioned above. In both equilibrium and
disequilibrum scenarios, for their nominal temperature profile and at solar
metallicity, a very narrow range of C/O ratios around 0.88 provides
the H$_2$O abundance constraint and a good fit to the observations. The recent retrieval analysis of Line et
al.\,(2013), using the same wavelength coverage of data, also finds a best-fit C/O ratio of 0.85
($\chi^{2}_{best}$/$N$=2.27 fit, with a 68\% confidence interval of
0.47-0.90). 

A carbon-enhanced atmosphere for HD 189733b is thus theoretically
plausible and consistent with observations. Interestingly, we find the
host star has C/O$=$0.90$\pm$0.15,  
matching well the
best-fit C/O ratios derived for the planet's atmosphere. HD 189733 is
the only star within this sample to have C/O$>$0.8; its C/O ratio
spans 0.75-1.05 within 1$\sigma$ errors. Three additional stars in our
sample have have C/O$>$0.8 within 1$\sigma$ errors. 

The derived T$_{\rm{eff}}$ of HD 189733 is $\leq$5400 K, therefore 
the triplet [O/H]$_{\rm{NLTE\,avg}}$ (0.125) is not included in the
final average [O/H] reported here. Instead, the triplet [O/H]$_{\rm{LTE}}$
(0.01$\pm$0.14) and [O/H]$_{\rm{6300}}$ ($-$0.02$\pm$0.14) values
are averaged.  
For stars as cool as HD 189733 there is
evidence from studies of [O/H] in several open clusters that the canonical
NLTE corrections are not appropriate -- [O/H]$_{\rm{LTE}}$
increases in lower-temperature stars in the same cluster,
the opposite of what is predicted  
(Schuler
et al.\,2006). If [O/H]$_{\rm{NLTE\,avg}}$ (0.125) is included in the
average, the C/O ratio of HD 189733 is reduced to
0.82, and if [O/H]$_{\rm{NLTE\,avg}}$ replaces
[O/H]$_{\rm{LTE}}$, the C/O ratio of HD 189733 is reduced to
0.79. The C/O ratio is reduced to 0.69 if
[O/H]$_{\rm{NLTE\,avg}}$ is the sole oxygen abundance
indicator. Alternatively, one may apply an empirical correction to
[O/H]$_{\rm{LTE}}$ based on the temperature of HD 189733 and the
observed cluster [O/H]$_{\rm{LTE}}$ anomaly (Schuler et al.\,2006),
which amounts to $\sim$0.14 dex. This increases the resulting C/O
ratio to 1.20. 

HD 189733's [C/H] is an
outlier as compared to the rest of our sample, while its [O/H] is more
consistent with the rest of the sample (Figure \ref{comparison_plot2}). Both measurements have some of the largest abundance errors of all the
targets in our sample. Our reported [C/H]$=$0.22$\pm$0.13 for HD
189733 is in fact based on only one carbon line, 5380\,{\AA}, from the
Keck/HIRES data, though we were able to measure two lines (5380 and
7111\,{\AA}) in the Subaru/HDS data, resulting in
[C/H]$=$0.24$\pm$0.15. 
Including the 7113\,{\AA}
C I line measurement from the
Keck/HIRES data or the Subaru/HDS data increases the [C/H]
from 0.22/0.24 to 0.34/0.33, and the reported C/O ratio to 1.16. Thus is
appears that the C/O ratio of HD 189733 could be as low as$\sim$0.75,
but is very likely $\gtrsim$0.80, as we report here. 

In order to match the desired H$_2$O
mixing ratio, the C/O ratio of the exoplanet HD 189733b's atmosphere must shift to
higher values when its metallicity is increased -- with an increase of 3$\times$solar in
metallicity the C/O ratio reaches $\sim$0.96, compared to our derived C/O
of 0.90$\pm$0.15. Alternatively, if the metallicity is sub-solar, the required C/O
ratio decreases (Moses et al.\,2013). Unfortunately with present data the metallicity
of HD 189733b's atmosphere is unknown. We find [Fe/H]=0.01$\pm$0.15 in
the host star, providing at least a first-order constraint on the
planetary atmospheric metallicity, but not a better constraint on its
C/O ratio. However, we note that based on Moses et al.\,(2013)'s
models, a change in the exoplanet's C/O ratio from 0.5 (solar) to
0.88 results in a change in the CH$_4$ abundance by $\sim$an order of
magnitude, which should produce observable spectral signatures in
the exoplanet's atmosphere.

\subsubsection{Future of Direct Planet-Star Comparisons}
The fact that HD 189733b is one of the most-studied hot
Jupiters and yet still has a C/O ratio that can be anywhere from $\sim$0.5-1
indicates the difficulty and uncertainty in deriving
exoplanet abundance ratios. Current limitations due larger to the
paucity of data, which gives rise to degenerate solutions for
transiting planet spectroscopy (e.g., Griffith 2013). However, as observational efforts continue to improve the
quantity and quality of the measurements more precise C/O ratios will
be possible. In addition, studies of transiting planets at high
spectral resolution are becoming progressively refined (e.g., Snellen
et al.\,2010; Birkby
et al.\,2013; de Kok et al.\,2013; Brogi et al.\,2013) to the
point that C/O ratio constraints are expected in the near future.

Complementary sutides of younger, hotter planets are possible with
spectroscopy of directly imaged planets. 
One such
system, HR 8799, is particularly promising as it has four
directly-imaged planets of similar luminosities, masses, and radii but
different orbital distances and, surprisingly, maybe even different
compositions (Barman et al.\,2011a; Currie et al.\,2011; Galicher et
al.\, 2011; Marley et al.\,2012; Skemer et al.\,2012, 2013; Konopacky et
al.\,2013).  
Recent directly-imaged,
moderate-resolution ($R\sim$4000) spectra from $\sim$1.97-2.38 $\mu$m
of HR 8799c show absorption of CO and H$_2$O but little to no CH$_4$
(mixing ratio $<$10$^{-5}$). $\chi^2$ minimization modeling of these
data finds best-fit log$N$(C) and log$N$(O) values of 8.33 and 8.51,
respectively, indicating  HR 8799c is depleted in both C and O with respect to solar, and resulting in a C/O ratio of 0.65$^{+0.10}_{-0.05}$ (Konoapacky et
al.\,2013). The host star is classified as both $\gamma$ Doradus and
$\lambda$ Bootis, making stellar abundance analysis challenging,
but one previous study of the star derives C/O=0.56$\pm$0.21
(Sadakane 2006). 
Determining the  
C/O ratios of the other planetary components in this system, and other
multi-planet systems, may provide  
constraints 
on how the composition of the host star affects giant planet
formation as a function of planet mass and orbital radius.  

\section{Summary}
The differences between [Fe/H] distributions in hosts versus
non-hosts have been the subject of study for over a decade, and a few more recent studies suggest that refractory
element distributions may differ in stars with/without planets, but
differences in volatile elements have not been as thoroughly
explored. Here we present a uniform stellar parameter and abundance analysis of
16 stars that host transiting hot Jupiter exoplanets. Our study also
includes one binary
companion that is not known to host planets. This work presents detailed measurements of transiting exoplanet host star carbon and oxygen
abundances, 
derived using muliple indicators of oxygen abundance. 
The derived host star C/O ratios
contribute one component to the direct comparison of stellar and
exoplanetary atmospheric compositions.

We compare our results to other studies of C/O vs. [Fe/H] in
exoplanet host stars, and find a similar positive slope between these
two parameters. This is indicative of Galactic chemical
evolution and the increasing importance of the carbon contribution
from the death of low- and intermediate-mass stars at more metal-rich
times in the Galaxy. A linear fit to our [C/H] versus [Fe/H] data
results in a slope $\sim$twice that of a linear fit to our [O/H]
vs. [Fe/H], with the former relation passing 0.08 dex below solar and
the latter passing 0.01 above solar. We derive a linear fit of
C/O=0.53[Fe/H] + 0.45 to our data, which falls 0.09 below our
adopted C/O$_{\odot}$=0.54, but corresponds to a 0.08 dex difference
in [C/O], similar to the offsets in [C/H] and [O/H] relations. These
offsets likely reflect both analysis uncertainties and the real,
intrinsic scatter in Galactic disk populations.

There is agreement between the average [C/H], [O/H], and C/O
values found here and the results of other studies of RV-detected
planet host stars, supporting previous findings (Ammler von-Eiff et
al.\,2009) that elemental abundance ratios do not differ significantly
between transiting and RV-planet host stars. The mean C/O ratio of the
transiting exoplanet host stars in this paper is slightly lower than
that found by other studies that consist of non-transiting host stars,
0.54$\pm$0.15 versus 0.71$\pm$0.07. This is more in line with recent
suggestions that the prior studies overestimated C/O ratios. 

Several
cases in which the process of directly comparing the chemistry in
specific stars to their planets is already beginning are highlighted
-- WASP-12, XO-1, TrES-2, TrES-3, HD 149026, XO-2, CoRoT-2, and HD 189733. We encourage
follow-up observational and theoretical studies of all of the
exoplanets whose host stars are included in this paper. 
Facilities that are currently available in
space and on the ground can be used strategically to obtain 
estimates of C/O ratios of a large sample of
transiting exoplanets, which JWST and several other upcoming
space-based missions (e.g., EChO) will be able to better
characterize. The more precise abundance analysis that is possible
right now for host stars can help infer their exoplanets'
formation histories, as well as inform future planet formation theories and models. 

\acknowledgements
The authors wish to recognize and acknowledge the very significant
cultural role and reverence that the summit of Mauna Kea has always
had within the indigenous Hawaiian community. We are most fortunate to
have the opportunity to conduct observations from this mountain. This
work would not have been possible without the efforts of the daytime
and nighttime support staff at the Mauna Kea Observatory and Subaru
Telescope, particularly Akito Tajitsu, Rita Morris, and Jennie
Berghuis. The authors also thank the anonymous referee for her/his
helpful corrections and comments. The work of J. T. and C. G. is
suppored by NASA's Planetary Atmospheres Program. J.T. thanks Andy Skemer for his aid in
interpreting directly-imaged planet results. This research has made
use of the Exoplanet Orbit Database and the Exoplanet Data Explorer at
exoplanets.org. 

{\it Facilities:} \facility{Subaru}, \facility{Keck}


\begin{deluxetable}{lccccc}
\tablecolumns{6}
\tablewidth{0pc}
\tabletypesize{\scriptsize}
\tablecaption{Observing Log \label{tab:log}}
\tablehead{  \colhead{Star} & \colhead{V} & \colhead{Date} &
  \colhead{Exposures} & \colhead{T$_{exp}$} & \colhead{Platform} \\
  \colhead{ } & \colhead{ } & \colhead{(UT)}& \colhead{ }&
  \colhead{(s)} & {}} 
\startdata
CoRoT-2   & 12.57 & 2013 Aug 30 & 3 & 1500 & Keck/HIRES \\ 
TrES-4    & 11.59 & 2013 Aug 30 & 2 & 1920, 900 & Keck/HIRES \\ 
TrES-2    & 11.25 & 2013 Aug 30 & 1 & 1320 & Keck/HIRES \\ 
WASP-2    & 11.98 & 2013 Aug 30 & 2 & 1440 & Keck/HIRES \\ 
WASP-12   & 11.57 & 2012 Feb 10 & 3 & 1080 (2), 1800 (1) &Subaru/HDS \\
          &       & 2012 Feb 11 & 2 & 1800 &Subaru/HDS\\
XO-2      & 11.25 & 2012 Feb 10 & 2 & 1800 &Subaru/HDS\\
XO-2B     & 11.20 & 2012 Feb 11 & 2 & 2100& Subaru/HDS\\
XO-1      & 11.25 & 2012 Feb 10 & 2 & 1800& Subaru/HDS\\
TrES-3    & 12.40 & 2012 Feb 10 & 2 & 2400& Subaru/HDS\\
          &       & 2012 Feb 11 & 2 & 2100& Subaru/HDS\\
          &       & 2008 Jun 12 & 17 &1200 (12), 600 (3), 420, 45 & Keck/HIRES archive \\ 
HD 189733 & 7.68  & 2012 Feb 10 & 2 & 120& Subaru/HDS\\
          &       & 2006 Aug 21 & 3 & 208, 212, 226 & Keck/HIRES archive \\ 
HD 149026 & 8.14  & 2012 Feb 11 & 1 & 480& Subaru/HDS\\
          &       & 2005 Jun 29 & 3 & 171,179,176 & Keck/HIRES archive \\ 
HD 80606  & 9.00  & 2012 Feb 11 & 2 & 600& Subaru/HDS\\
HAT-P-7   & 10.48 & 2012 Feb 10 & 1 & 900& Subaru/HDS\\
          &       & 2012 Feb 11 & 2 & 1200& Subaru/HDS\\
          &       & 2007 Aug 24 & 1 & 600 & Keck/HIRES archive \\
HAT-P-13  & 10.42 & 2012 Feb 11 & 1 & 1800& Subaru/HDS\\
HAT-P-1   & 10.4  & 2012 Aug 31 & 2 & 500, 630 & Keck/HIRES\\
HAT-P-16  & 10.91 & 2012 Aug 31 & 2 & 1120, 1043 & Keck/HIRES \\
WASP-32   & 11.26 & 2012 Aug 31 & 2 & 1500, 1800 & Keck/HIRES\\
Moon      &       & 2012 Feb 10 & 2 & 1, 5 &Subaru/HDS\\
Vesta     &       & 2006 Apr 16 & 3 & 216, 232, 241 & Keck/HIRES archive\\
\enddata
\end{deluxetable}

\begin{landscape}
\begin{deluxetable}{lccccc}
\tablecolumns{6}
\tablewidth{0pc}
\tabletypesize{\scriptsize}
\tablecaption{Observing Platform Details \label{tab:platform}}
\tablehead{  \colhead{Platform} & \colhead{slit} & \colhead{$R$} &
  \colhead{Wavelength Coverage} & \colhead{S/N of combined frames}  & \colhead{Seeing Range} \\
  \colhead{ } & \colhead{(and filter, if applicable)} & \colhead{($
    \frac{\lambda}{\Delta\lambda}$)} &\colhead{(\AA)}& \colhead{(at
    6300\,\AA)} &\colhead { } }
\startdata
Subaru/HDS & 0.''6 & 60,000 & $\sim$4450-5660; 5860-7100$^{*}$ &
$\sim$170-230 & 0.''84-1.''12; 0.''96-1.''24$^{+}$ \\
Keck/HIRES & 0.''86 (C1 decker) & 48,000 & $\sim$3360-8100 &
$\sim$125-150 & $\sim$0.4-0.6''; $\sim$0.6-0.8''$^{++}$ \\
           & kv370+clear filters & & & & \\
Keck/HIRES archive & & & & & \\
HAT-P-7& same & same & same &   $\sim$190 & \\
HD 189733& same & same & same & $\sim$250 & \\
HD 149026& same & same & same & $\sim$290&  \\
Vesta& same & same & same&  $\sim$315 & \\
TrES-3&0.''57, kv389 filter  &72,000 &4240-8690 &$\sim$300   &  \\
\enddata
\tablenotetext{*}{Wavelength coverage across two separate CCDs.}
\tablenotetext{+}{Seeing from Feb. 10; Feb. 11.}
\tablenotetext{++}{Seeing from Aug. 31, 2012; Aug. 30, 2013}
\tablecomments{We note that the TrES-3 Keck/HIRES archive observations used a different
  filter and narrower slit, and thus had slightly different wavelength
  coverage and higher resolution, but this did not affect our ability
  to measure the necessary elemental absorption lines.}
\end{deluxetable}
\end{landscape}

\begin{deluxetable}{lllllllllllll}
\tabletypesize{\scriptsize}
\tablecolumns{13}
\tablewidth{0pc}
\tablecaption{Derived Stellar Parameters \label{tab:stellar_params}}
\tablehead{   \colhead{Star} & \colhead{T$_{eff}$} & \colhead{$\sigma$} &
  \colhead{log $g$} & \colhead{$\sigma$} & \colhead{$\xi$} &
  \colhead{$\sigma$} & \colhead{[Fe I/H]} & \colhead{$N$} &
  \colhead{$\sigma_{\mu}$} &  \colhead{[Fe II/H]} & \colhead{$N$} &
  \colhead{$\sigma_{\mu}$} \\ \colhead{ } & \colhead{K} & \colhead{K} &  \colhead{ } & \colhead{ } & \colhead{km s$^{-1}$} &
  \colhead{km s$^{-1}$} & \colhead{} & \colhead{ } & \colhead{ } &
  \colhead{ } & \colhead{ } & \colhead{ } }

\startdata

CoRot-2  & 5616 & 47 &  4.52  &  0.14 & 1.59 & 0.09 & 0.063 & 48 & 0.007 & 0.064 & 7 & 0.017 \\
TrES-4 & 6333 &  44 &  4.04  & 0.17 & 1.74 & 0.09 & 0.320 & 49  & 0.005 & 0.322 & 9 & 0.025 \\
TrES-2  & 5823 &  33 &  4.45  & 0.10 & 1.27 & 0.07 & -0.016 & 51 & 0.004 & -0.016 & 9 & 0.005 \\
WASP-2  & 5228 &  60 &  4.49  & 0.21 & 1.07 & 0.10 & 0.092 & 52 & 0.008 & 0.091 & 9 & 0.021 \\
WASP-12   & 6166 & 41 & 4.05 & 0.16 & 1.95 & 0.13 & 0.062 & 40 & 0.006&0.062 & 10 & 0.020 \\ 
XO-2N      & 5343 & 78 & 4.49 & 0.25 & 1.22 & 0.09 & 0.386 & 49 &0.011& 0.389 & 8 & 0.020 \\
XO-2S     & 5547 & 59 & 4.22 & 0.24 & 1.24 & 0.07 & 0.291 & 50 &0.010&0.295 & 10 & 0.048 \\
XO-1      & 5695 & 26 & 4.42 & 0.12 & 1.39 & 0.06 & $-$0.109 & 36 & 0.004 &$-$0.110 &9 &0.008 \\
TrES-3    & 5534 & 42 & 4.56 & 0.14 & 1.20 & 0.10 & $-$0.209 & 33 & 0.007 & $-$0.206 & 8 & 0.021  \\
HD 189733 & 5116 & 76 & 4.64 & 0.25 & 1.27 & 0.16 & 0.012 & 43 &0.012 &0.011 & 9 & 0.041 \\ 
HD 149026 & 6093 & 48 & 4.30 & 0.21 & 1.71 & 0.09 & 0.265 & 51 & 0.007& 0.264 & 9 & 0.024 \\
HD 80606  & 5551 & 47 & 4.14 & 0.17 & 1.29 & 0.06 & 0.274 & 41 & 0.008& 0.275 & 8 &0.046 \\
HAT-P-7   & 6474 & 71 & 4.33 & 0.29 & 2.72 & 0.37 & 0.140 & 40 & 0.008& 0.139 & 10 & 0.033 \\ 
HAT-P-13  & 5775 & 57 & 4.13 & 0.17 & 1.44 & 0.07 & 0.442 & 51 & 0.009& 0.445 & 10 & 0.035 \\
HAT-P-1   & 6045 & 44 & 4.52 & 0.12 & 1.51 & 0.11 & 0.172 & 53
&0.006&0.174   & 8  &0.012 \\
HAT-P-16  & 6236 & 58 & 4.49 & 0.19 & 1.58 & 0.15 & 0.174 & 54 & 0.007& 0.172 & 9 & 0.015 \\
WASP-32   & 6042& 42&4.34&0.20&1.80&0.15&$-$0.066&53&0.006 &$-$0.069 &9 & 0.023 \\ 

\enddata
\end{deluxetable}
\clearpage

\begin{landscape}
\begin{deluxetable}{llllllllllllll}
\tablecolumns{14}
\tablewidth{0pc}
\tabletypesize{\scriptsize}
\tablecaption{Lines Measured, Equivalent Widths, and Abundances \label{tab:lines}}
\tablehead{ \colhead{Ion} & \colhead{$\lambda$} & \colhead{$\chi$} &
  \colhead{log $gf$} & \colhead{EW$_{\odot}$} &
  \colhead{log$N_{\odot}$} & \multicolumn{2}{c}{WASP-12} & \colhead{ }
  & \multicolumn{2}{c}{HD 149026} & \colhead{ } & \multicolumn{2}{c}{HAT-P-1} \\ 
  \cline{7-8} \cline{10-11} \cline{13-14} \\
  \colhead{ } & \colhead{(\AA)} & \colhead{(eV)} & \colhead{(dex)} &
  \colhead{(m\AA)} & \colhead{ } & \colhead{EW (m\AA)} & \colhead{log
    $N$} & \colhead{ } & \colhead{EW (m\AA)} & \colhead{log $N$} &
  \colhead{ } & \colhead{EW (m\AA)} & \colhead{log $N$} \\}

\startdata
C I & 5052.17 & 7.68 & -1.304 & 33.9$^{a}$, 33.7$^{b}$& 8.46$^{a}$,
8.45$^{b}$& 62.4$^{a}$ & 8.58$^{a}$ & \nodata & \nodata & \nodata &
\nodata & 43.9$^{b}$ & 8.51$^{b}$ \\
{ } & 5380.34 & 7.68 & -1.615 & 19.4$^{a}$, 20.7$^{b}$ & 8.44$^{a}$, 8.48$^{b}$ & 41.1$^{a}$ & 8.57$^{a}$ & \nodata & 44.5$^{b}$ & 8.76$^{b}$ & \nodata & 29.3$^{b}$ & 8.56$^{b}$ \\ 
{ } & 6587.61 & 8.54 & -1.021 & 12.9$^{a}$, 15.5$^{b}$ & 8.38$^{a}$, 8.48$^{b}$ & 27.0$^{a}$ & 8.4$^{a}$ & \nodata & 36.4$^{b}$ & 8.75$^{b}$ & \nodata & 20.2$^{b}$ & 8.46$^{b}$ \\
{ } & 7111.47 & 8.64 & -1.074 & 9.8$^{a}$, 12.2$^{b}$  & 8.38$^{a}$, 8.50$^{b}$ & 21.0$^{a}$ & 8.42$^{a}$ & \nodata & 26.6$^{b}$ & 8.69$^{b}$ & \nodata & 16.3$^{b}$ & 8.49$^{b}$ \\
{ } & 7113.18 & 8.65 & -0.762 & 22.8$^{a}$, 20.9$^{b}$ & 8.55$^{a}$, 8.50$^{b}$ & \nodata & \nodata & \nodata & 47.4$^{b}$ & 8.79$^{b}$ & \nodata & 29.2$^{b}$ & 8.53$^{b}$ \\
$\rm{[O I]}^{*}$ & 6300.30 & 0.00 & -9.717 & 5.4$^{a}$, 5.6$^{b}$&  8.68$^{a}$, 8.67$^{b}$ & 7.0$^{a}$ & 8.82$^{a}$ & \nodata & 8.2$^{b}$ & 8.95$^{b}$ & \nodata & 4.6$^{b}$ & 8.62$^{b}$ \\
O I  & 7771.94 &  9.15&  0.37  & 69.6$^{b}$ & 8.83$^{b}$ &\nodata & \nodata & \nodata & 119.8$^{b}$ & 9.17$^{b}$ & \nodata & 89.7$^{b}$ & 8.88$^{b}$\\
O I  & 7774.17 & 9.15 & 0.22  & 62.6$^{b}$&8.86$^{b}$&\nodata &
\nodata & \nodata & 108.6$^{b}$ & 9.18$^{b}$ & \nodata & \nodata &\nodata \\
O I  & 7775.39 & 9.15 &  0.00 &  46.8$^{b}$          & 8.81$^{b}$&\nodata & \nodata & \nodata &82.0$^{b}$ & 9.05$^{b}$ &  & 61.1$^{b}$ &8.83$^{b}$ \\

\enddata
\tablenotetext{a}{Measured in the Subaru/HDS data.}
\tablenotetext{b}{Measured in the Keck/HIRES data.}
\tablenotetext{*}{For [O I], the log $N$ values represent those
  derived from synthesis fitting, as these are the values we use in
  calculating the final [O/H] for each object. The reported EWs refer
  to the total EW of the 6300.3\,{\AA} blend.}
\tablecomments{The full version of this table including all lines and targets is available online.}
\end{deluxetable}
\end{landscape}

\begin{landscape}
\begin{deluxetable}{lccccc}
\tablecolumns{6}
\tabletypesize{\scriptsize}
\tablewidth{0pc}
\tablecaption{Oxygen Abundances Derived from Different Indicators \label{tab:oh}}
\tablehead{ 
\colhead{Star} & \colhead{[O/H]} & \colhead{[O/H]} & \colhead{[O/H]} & \colhead{[O/H]} & \colhead{[O/H]} \\
\colhead{ } & \colhead{[O I] 6300\,{\AA}} & \colhead{triplet LTE} & \colhead{triplet NLTE, Takeda} & 
\colhead{triplet NLTE, Ram{\'{\i}}rez} & \colhead{triplet NLTE, Fabbian} }
\startdata
CoRoT-2$^{*}$                  & \nodata &0.02$\pm$0.07& 0.07& 0.06 & 0.07 \\ 
TrES-4$^{*}$                &0.22$\pm$0.09&0.31$\pm$0.07&0.15 & 0.21 &0.08 \\
TrES-2$^{*}$                 &-0.02$\pm$0.05&0.00$\pm$0.05&0.02&0.01&-0.01\\
WASP-2$^{*}$                   &-0.03$\pm$0.10&-0.01$\pm$0.10&0.08&0.06&0.09\\
WASP-12                   & 0.14$\pm$0.06    & \nodata          &\nodata    & \nodata & \nodata   \\
XO-2N                     & 0.34$\pm$0.16    & \nodata          &\nodata    & \nodata & \nodata   \\
XO-2S                    & 0.18$\pm$0.15    & \nodata          &\nodata    & \nodata & \nodata   \\
XO-1                    & $-$0.09$\pm$0.05    & \nodata          &\nodata    & \nodata & \nodata   \\
TrES-3$^{*}$            & $-$0.04$\pm$0.06 & \nodata & \nodata &\nodata \\      
HD 189733$^{*}$         & $-$0.02$\pm$0.14 & 0.01$\pm$0.14 & 0.12 &0.11 & 0.14 \\
HD 149026$^{*}$            & 0.28$\pm$0.03  & 0.30$\pm$0.07  & 0.21 & 0.26   & 0.20\\
HD 80606                   & 0.20$\pm$0.08   & \nodata          &\nodata    & \nodata & \nodata   \\
HAT-P-7$^{*}$             & \nodata        & 0.22$\pm$0.10 & 0.06 & 0.11 & 0.02\\
HAT-P-13                   & 0.19$\pm$0.08   & \nodata          &\nodata    & \nodata & \nodata   \\
HAT-P-1$^{*}$              &$-$0.05$\pm$0.06 &  0.07$\pm$0.06 & 0.07 & 0.07 & 0.02\\
HAT-P-16$^{*}$               &$-$0.08$\pm$0.10 & 0.04$\pm$0.06 & 0.03& $-$0.01 &$-$0.10 \\
WASP-32$^{*}$               &$-$0.08$\pm$0.09 & 0.08$\pm$0.08 & 0.03 & 0.00 & $-$0.01 \\
\enddata
\tablenotetext{*}{Measurements from Keck/HIRES data.}
\end{deluxetable}
\end{landscape}

\clearpage

\begin{landscape}
\begin{deluxetable}{lccccc}
\tablecolumns{6}
\tabletypesize{\scriptsize}
\tablewidth{0pc}
\tablecaption{Elemental Abundances and Ratios \label{tab:compare}}
\tablehead{ 
\colhead{Star} & \colhead{[Fe/H]} & \colhead{[C/H]} & \colhead{[O/H]$_{\rm{avg}}$} &
  \colhead{[Ni/H]} & \colhead{C/O$_{\rm{avg}}$}}

\startdata
CoRoT-2  & 0.06$\pm$0.08$^{*}$  & 0.01$\pm$0.06$^{*}$   & 0.06$\pm$0.07$^{*}$     & -0.08$\pm$0.03$^{*}$   &0.47$\pm$0.09$^{*}$   \\
TrES-4  & 0.32$\pm$0.09$^{*}$  & 0.11$\pm$0.06$^{*}$   & 0.18$\pm$0.06$^{*}$     & 0.29$\pm$0.02$^{*}$   &0.46$\pm$0.08$^{*}$   \\
TrES-2  & -0.02$\pm$0.05$^{*}$  &-0.12$\pm$0.04$^{*}$   & -0.01$\pm$0.04$^{*}$     & -0.08$\pm$0.02$^{*}$   &0.41$\pm$0.05 $^{*}$  \\
WASP-2  & 0.09$\pm$0.12$^{*}$  & -0.01$\pm$0.09$^{*}$   & -0.02$\pm$0.07$^{*}$     & 0.11$\pm$0.03$^{*}$   &0.55$\pm$0.11$^{*}$   \\
WASP-12  & 0.06$\pm$0.08  & 0.09$\pm$0.06   & 0.14$\pm$0.06     & 0.00$\pm$0.04   &0.48$\pm$0.08   \\
XO-2N   & 0.39$\pm$0.14   & 0.42$\pm$0.12   &0.34$\pm$0.16  & 0.44$\pm$0.04    & 0.65$\pm$0.20\\
XO-2S   & 0.28$\pm$0.14  & 0.26$\pm$0.11   & 0.18$\pm$0.15    & 0.38$\pm$0.04    & 0.65$\pm$0.19\\
XO-1    & $-$0.11$\pm$0.06 & $-$0.19$\pm$0.04& $-$0.09$\pm$0.05  &$-$0.11$\pm$0.02  & 0.43$\pm$0.07\\
TrES-3 & $-$0.21$\pm$0.08 & $-$0.31$\pm$0.06$^{*}$&  $-$0.04$\pm$0.06$^{*}$ &$-$0.25$\pm$0.04 & 0.29$\pm$0.09$^{*}$\\
HD 189733 & 0.01$\pm$0.15 & 0.22$\pm$0.11$^{*}$   & -0.01$\pm$0.10$^{*}$ &0.00$\pm$0.05 & 0.90$\pm$0.15$^{*}$\\
HD 149026 & 0.26$\pm$0.09 & 0.26$\pm$0.08$^{*}$   & 0.25$\pm$0.04$^{*}$  & 0.31$\pm$0.03    & 0.55$\pm$0.08$^{*}$\\
HD 80606  & 0.28$\pm$0.10  & 0.29$\pm$0.08   & 0.20$\pm$0.08     & 0.30$\pm$0.03    & 0.66$\pm$0.12\\
HAT-P-7  & 0.14$\pm$0.14 & $-$0.04 $\pm$0.10$^{*}$  & 0.07$\pm$0.10$^{*}$  & 0.12$\pm$0.05  & 0.42$\pm$0.14$^{*}$\\
HAT-P-13  & 0.44$\pm$0.09 & 0.34$\pm$0.08   & 0.19$\pm$0.08      & 0.53$\pm$0.04    & 0.76$\pm$0.11  \\
HAT-P-1  &0.17$\pm$0.06$^{*}$ &0.03$\pm$0.05$^{*}$ & 0.00$\pm$0.04$^{*}$ & 0.17$\pm$0.03$^{*}$ & 0.58$\pm$0.06$^{*}$\\
HAT-P-16 &0.17$\pm$0.09$^{*}$ & $-$0.02$\pm$0.06$^{*}$ & $-$0.05$\pm$0.06$^{*}$ & 0.13$\pm$0.04$^{*}$ &0.58$\pm$0.08$^{*}$ \\  
WASP-32  &$-$0.07$\pm$0.09$^{*}$ & $-$0.09$\pm$0.07$^{*}$ & $-$0.04$\pm$0.06$^{*}$ & $-$0.13$\pm$0.03$^{*}$ &0.47$\pm$0.09$^{*}$ \\ 
\enddata

\tablecomments{C/O=10$^{logN(C)}$/10$^{logN(O)}$, with
  log$N$(C)$=$derived\,[C/H]+log$N_{\odot}$(C) and log$N$(O)$=$derived\,[O/H]+log$N_{\odot}$(O),
where log$N_{\odot}$(O)$=$8.66 and log$N_{\odot}$(C)$=$8.39 (solar
values from Asplund et al.\,2005). The errors on the C/O ratio are represented by the
quadratic sum of the errors in [C/H] and [O/H].}
\tablenotetext{*}{Measurements include Keck/HIRES data.}
\end{deluxetable}
\end{landscape}

\begin{deluxetable}{lccccccc}
\tablecolumns{8}
\tablecaption{Abundance Sensitivities \label{tab:sens}}
\tabletypesize{\scriptsize}
\tablewidth{0pc}
\tablehead{
  \colhead{Species} & \multicolumn{3}{c}{WASP-12} & \colhead{ } & \multicolumn{3}{c}{HAT-P-1} \\ 
\cline{2-4} \cline{6-8} \\
\colhead{ } & \colhead{$\Delta$T$_{eff}$} & \colhead{$\Delta$log $g$} &
\colhead{$\Delta \xi$} & \colhead{ } & \colhead{$\Delta$T$_{eff}$} & \colhead{$\Delta$log $g$} &
\colhead{$\Delta \xi$} \\
\colhead{ } & \colhead{($\pm$150 K)} & \colhead{($\pm$0.25 dex)} &
\colhead{($\pm$0.30 km s$^{-1}$)} & \colhead{ } & \colhead{($\pm$150 K)} & \colhead{($\pm$0.25 dex)} &
\colhead{($\pm$0.30 km s$^{-1}$)}}
\startdata
Fe I  & $\pm$0.10 & $\pm$0.005 & $\pm$0.03 &  & $\pm$0.09 & $\pm$0.005 & $\pm$0.03 \\
Fe II & $\pm$0.02 & $\pm$0.10 & $\pm$0.08 &  & $\pm$0.05 & $\pm$0.10 & $\pm$0.07\\
C I   & $\pm$0.08 & $\pm$0.08 & $\pm$0.01 &  & $\pm$0.10 & $\pm$0.08 & $\pm$0.005 \\
$\rm{[O I]}^{*}$ & $\pm$0.06 & $\pm$0.09 & $\pm$0.01 &  & $\pm$0.11 & $\pm$0.12 & $\pm$0.00 \\
Ni I  & $\pm$0.11 & $\pm$0.02 & $\pm$0.03 &  & $\pm$0.10 & $\pm$0.005 & $\pm$0.04 \\
O I triplet (LTE) & \nodata & \nodata & \nodata & & $\pm$0.13 &
$\pm$0.07 & $\pm$0.03\\
\enddata
\tablenotetext{*}{For [O I], the log $N$ values represent those
  derived from synthesis fitting, as these are the values we use in
  calculating the final [O/H] for each object.}
\end{deluxetable}

\clearpage
\begin{landscape}
\begin{deluxetable}{lcccccc}
\tablecolumns{7}
\tabletypesize{\scriptsize}
\tablewidth{0pt}
\tablecaption{Comparison of Average C \& O Measurements to Previous Work \label{tab:compare2}}
\tablehead{
\colhead{Source}                                &
\colhead{$\overline{[C/H]}_{\rm{hosts}}$} & \colhead{$\overline{[C/H]}_{\rm{non-hosts}}$} &
\colhead{$\overline{[O/H]}_{\rm{hosts}}$} & \colhead{$\overline{[O/H]}_{\rm{non-hosts}}$} &
\colhead{$\overline{C/O}_{\rm{hosts}}$} & \colhead{$\overline{C/O}_{\rm{non-hosts}}$} \\ }
\startdata
Ecuvillon et al. \,(2004) or (2006) &  0.14 $\pm$ 0.10 & $-$0.03
$\pm$0.14 & 0.12 $\pm$ 0.11 & 0.07 $\pm$ 0.15 & \nodata & \nodata \\    
Bond et al.\,(2006) or (2008) & 0.17 $\pm$ 0.11 & 0.01 $\pm$ 0.17 &
0.00 $\pm$ 0.17 & $-$0.06 $\pm$0.15 & 0.67 $\pm$ 0.23 & 0.67 $\pm$
0.23 \\
Delgado Mena et al.\,(2010) & 0.10 $\pm$ 0.16 & $-$0.06 $\pm$ 0.18 &
0.05 $\pm$ 0.17 & $-$0.08 $\pm$ 0.17 & 0.76 $\pm$ 0.20 & 0.71 $\pm$
0.18 \\
Petigura \& Marcy\,(2011) & 0.17 $\pm$ 0.14 & 0.08 $\pm$ 0.17 & 0.11
$\pm$ 0.12 & 0.05 $\pm$ 0.14 & 0.76 $\pm$ 0.22 & 0.70 $\pm$ 0.22 \\
Nissen\,(2013) & 0.11 $\pm$ 0.15 & \nodata & 0.08 $\pm$ 0.10 & \nodata
& 0.63 $\pm$ 0.12 & \nodata \\ 
this work (\textit{only transiting planets}) &0.06$\pm$0.20 & \nodata & 0.07$\pm$0.13 & \nodata & 0.54$\pm$0.15& \nodata \\
\enddata
\tablecomments{$\rm{L}$isted are the means and standard deviations in
  exoplanet hosts stars and ``non-host'' stars, for each elemental abundance
  ratio, given as (mean $\pm$ standard deviation). Note that the
  number of objects in each source's sample is not equal, and that
  different sources use different solar log$N$(C) and log$N$(O) values.} 
\end{deluxetable}

\clearpage
\end{landscape}


\end{document}